\documentclass{aa}  

\usepackage{graphicx}
\usepackage{xcolor}
\usepackage{txfonts}
\usepackage{natbib}
\begin{document}

\title{Exploring the ultra-faint dwarf Bo\"otes I using JWST and HST: Metallicity distribution and binaries}

   \author{F. Muratore \inst{1}, 
       M. V. Legnardi \inst{1}, 
        A. P. Milone \inst{1,2}, 
        A. Mastrobuono-Battisti \inst{1,2,3},
        G. Cordoni \inst{4}, 
        L. N. Gorza \inst{1}, 
        E. P. Lagioia \inst{5}, 
        E. Bortolan \inst{1}, 
        E. Dondoglio \inst{2}, 
        A. F. Marino \inst{2}, 
        T. Ziliotto \inst{1} 
   }

   \institute{Dipartimento di Fisica e Astronomia “Galileo Galilei”, Università Degli Studi di Padova, Vicolo dell’Osservatorio 3, 35122 Padova, Italia
     \and
     Istituto Nazionale di Astrofisica - Osservatorio Astronomico di Padova, Vicolo dell’Osservatorio 5, 35122 Padova, Italy 
     \and
    Dipartimento di Tecnica e Gestione dei Sistemi Industriali, Università degli Studi di Padova, Stradella S. Nicola 3, I-36100 Vicenza, Italy  
     \and 
     Research School of Astronomy and Astrophysics, Australian National University, Canberra, ACT 2611, Australia  
      \and
    South-Western Institute for Astronomy Research, Yunnan University, Kunming 650500, PR China 
}

   \date{Received May XX, XXXX; accepted May XX, XXXX}
\titlerunning{Binaries and metallicity distribution in Bo\"otes I} 
\authorrunning{Muratore et al.}


\abstract
{ 
Ultra-faint dwarf galaxies (UFDs) are among the oldest and most metal-poor stellar systems in the Universe. Their metallicity distribution encodes the fossil record of the earliest star formation, feedback, and chemical enrichment, providing crucial tests of models of the first stars, galaxy assembly, and dark matter halos. However, due to their faint luminosities and the limited number of bright giants, spectroscopic studies of UFDs typically probe only small stellar samples.
Here, we present an analysis of multi-epoch \textit{Hubble} Space Telescope and \textit{James Webb} Space Telescope observations of the UFD Bo\"otes I. Using deep color–magnitude diagram in the F606W and F322W2 bands, extending from the subgiant branch to the M-dwarfs, and stellar proper motions to identify likely members, we obtained an unprecedentedly clean census of the system. The exquisite quality of the diagram, combined with the sensitivity of M-dwarf colors to metallicity, allowed us to constrain the metallicity distribution in a large stellar sample.
As a first step, we then exploited the metallicity sensitivity of M-dwarf colors to derive the metallicity distribution function. We find that most of the stars $\sim$85\% have [Fe/H] $< -2$, and that roughly $\sim$17\% have [Fe/H] $< -3$. Then, we derived the binary fraction in Bo\"otes I. This is crucial, since binaries can bias kinematic mass estimates, affect stellar population analyses, and shape the photometric signatures used to infer metallicity. We find that 20$\pm2$\% of stellar systems in Bo\"otes I are binaries with mass ratios larger than 0.4, corresponding to a total binary fraction of $\sim$30\%.  This value is comparable to the binary fractions observed in globular clusters of similar stellar mass, suggesting that the presence of dark matter does not significantly affect the binary properties of Bo\"otes I. 
}

   \keywords{dwarf galaxy -- Stellar populations -- Binary stars -- metallicity distribution}

   \maketitle
%

\section{Introduction}

Ultra-faint dwarf galaxies (UFDs) are among the oldest and most chemically primitive stellar systems known, typically defined as having total luminosities below 10$^{5}$L$_\odot$ \citep[e.g.\,][and references therein]{simon2019}. 
Over the past two decades, numerous UFDs have been identified through wide-field imaging surveys \citep[e.g.][]{belokurov2010a, willman2011, kim2015a}. Although they may superficially resemble globular clusters (GCs), UFDs display distinctive structural and dynamical properties that clearly distinguish them as separate systems.

One of the most intriguing features of UFDs is their extreme dark matter content: their total dynamical masses exceed the stellar mass by factors of 10$^{2}$–10$^{4}$, corresponding to mass-to-light ratios of similar magnitude \citep[e.g.][]{simon2007,mcconnachie2012, battaglia2022}. Owing to their relative proximity, UFDs represent unique laboratories for probing the nature and distribution of dark matter on the smallest galactic scales.
Moreover, owing to their extremely low baryonic content and their residence in low-mass, dark-matter–dominated halos, UFDs represent ideal laboratories for constraining the nature of dark matter \citep{Zoutendijk2021a} and for testing models proposed to explain the missing satellite problem \citep{simon2019}. Furthermore, deep \textit{Hubble Space} Telescope  (HST) observations have shown that the stellar populations of UFDs are predominantly ancient, supporting the view that they are “fossil” galaxies—systems that formed most of their stars before reionization and subsequently ceased forming stars \citep[e.g.][]{brown2014, sacchi2021, savino2023, Durbin2025}.

Reliable estimates of the dark matter content in UFDs are currently obtained from measurements of the radial–velocity dispersion of their member stars. Early studies already indicated that UFDs contain a large amount of dark matter based on their unexpectedly high velocity dispersions. 
More recent works on Bo\"otes~I and Leo~IV reported dispersions of \(4-5 \,\,\mathrm{km\,s^{-1}}\) and \(\sim 3\,\mathrm{km\,s^{-1}}\), respectively  supporting the presence of a dark matter halo \citep[e.g.][]{jenkins2021,longeard2022,sandford2025}.

However, unresolved binary systems can artificially inflate the observed velocity dispersion, leading to overestimates of the inferred dark matter mass \citep{spencer2018,pianta2022}. Consequently, constraining the binary fraction and the properties of binary stars is essential for obtaining robust dynamical measurements \citep{spencer2017,spencer2018,mcconnachie2010, gration2025}.

Another intriguing characteristic of UFDs is the considerable variation in iron (Fe) and $\alpha$-element abundances observed among their stars, interpreted as the result of extended star formation episodes and internal chemical enrichment \citep{simon2007, kirby2008}. These chemical patterns contrast sharply with those seen in GCs, which generally display much smaller star-to-star metallicity variations \citep[e.g.][]{marino2015,legnardi2022}, resembling those observed in brighter dwarf galaxies. At the same time, UFDs are less complex than massive galaxies, making them ideal laboratories for studying the fundamental mechanisms that govern chemical evolution in low-mass galaxies.
However, metallicity determinations mostly rely on spectroscopy of red giant branch (RGB) stars, which are typically few in number \citep{norris2010a,lai2011,frebel2016}. This limitation arises from the sensitivity of current telescopes, which cannot obtain high-resolution spectra for faint stars, combined with the sparsely populated RGBs characteristic of most UFDs.

In this paper, we take advantage of imaging data collected with the HST and the \textit{James Webb} Space Telescope (JWST) to study Bo\"otes\,I, one of the most extensively investigated UFDs. The resulting high-precision photometry enables us to pursue two main goals: (i) the determination of the metallicity distribution, and (ii) the characterization of the binary fraction along the main sequence (MS). 

Discovered by \citet{belokurov2006}, Bo\"otes\,I has since been the subject of numerous investigations of its stellar populations and chemical composition. With a luminosity of $L_V = 2.8 \times 10^4,L_{\odot}$, Bo\"otes\,I is among the brightest known UFDs. Using the Ca\textsc{ii} K line–strength index, \citet{norris2008} reported a mean [Fe/H] of $-2.51 \pm 0.13$ for a sample of 16 stars. Further high-resolution spectroscopic studies have confirmed a mean value of $\langle\mathrm{[Fe/H]}\rangle \simeq -2.6$, a wide iron range of $\sim$2 dex, and very metal-poor stars with [Fe/H] $<-3.5$ \citep{norris2010b, koposov2011, gilmore2013, ishigaki2014}. Similar results were also inferred using low- to medium-resolution spectroscopy \citep{lai2011, jenkins2021, longeard2022, sandford2025}. As noted above, Bo\"otes\,I exhibits a relatively large line-of-sight velocity dispersion, implying a high mass-to-light ratio and a significant dark-matter component.

Bo\"otes\,I is composed of old stellar populations and hosts a conspicuous fraction of binaries. From an analysis of the color-magnitude diagram (CMD) based on HST data, \citet[][see also \citealt{Durbin2025}]{brown2014} inferred an age of about 13\,Gyr, while \citet{gennaro2018} used the same data to constrain a binary fraction of $\sim0.28$ to characterize the mass function. The main structural and physical properties of Bo\"otes\,I are summarized in Table~\ref{tab1}.

The paper is organized as follows. Section\,\ref{sec2} describes the dataset and the methods for deriving high-precision photometry and astrometry. 
In Sect.\,\ref{sec_cmd}, we present the CMD  of Bo\"otes\,I.
Section \ref{sec_met} describes the methods to infer the metallicity distribution of stars in Bo\"otes\,I by using photometry of M-dwarfs, while Sect.\,\ref{sec_bin} is dedicated to the determination of the binary fraction and the investigation of the binary stars in Bo\"otes\,I. Finally, Sect.\,\ref{sec_con} provides a summary and a discussion of the results.

\begin{table}[h]
    \centering
        \caption{Main observational and physical parameters of Bo\"otes I.}
    \begin{tabular}{l l l}
        \hline
        Parameter & Value & Reference \\
        \hline
        RA(J2000) & $14^h00^m06^s$ & 1 \\
        Dec.(J2000) & +$14^{\circ}30'00''$  & 1 \\
        $M_V$ & $-5.92$  $\pm$ $0.2$ mag & 2 \\
        D & 65  $\pm$ 3 kpc & 2 \\
        $r_h$ & $12.5$  $\pm$ $0.3$ arcmin & 2 \\
        $r_t$ & 33.1  $\pm$ 4.0 arcmin & 3 \\
        \textnormal{[Fe/H]} & $-2.43$  $\pm$ $0.4$ &\textbf{ 4} \\
        $M_* $ &  $3.4 \pm 0.3 \times 10^4 M_{\odot}$    &   5  \\
        $M_{dyn} $ &   $4.9^{+1.3}_{-1.2} \times 10^6 M_{\odot}$  &   6  \\        
        \hline
    \end{tabular}
    References:(1) \cite{belokurov2006}, (2) \cite{okamoto2012}, (3) \cite{koposov2011}, (4) \cite{sandford2025}, (5) \cite{martin2008}, (6) \cite{jenkins2021}
    \label{tab1}
\end{table}

\section{Data}
\label{sec2}
The data for this study were obtained from  HST GO 12549 (PI: Brown), GO 15317 (PI: Platais), and JWST GO 3849 (PI: Gennaro).

\begin{table*}[h]
    \centering
        \caption{Summary of the HST and JWST imaging data used in this study. Since the three observed fields share the same number of exposures and exposure times, we report these values only once. }
    \begin{tabular}{l l l l l l l}
        \hline
         Filter & Instrument  &Camera& N x Exposure time & Date & Program & PI\\
        \hline
         F606W & HST & WFC/ACS  & 2 x 500 + 2 x 670 s & May 25 - June 13, 2012 & 12549 & Brown \\
         F814W & HST & WFC/ACS  & 2 x 430 + 2 x 670 s & May 25 - June 13, 2012 & 12549 & Brown \\
         F606W & HST & WFC/ACS  & 8 x 1224 s & June 10 - 29,  2019 & 15317 & Platais \\
         F814W & HST & WFC/ACS  & 8 x 1258 s & June 10 - 29, 2019 & 15317 & Platais \\
         F322W2 & JWST & NIRCam &  20 x 945 s & July 6, 2024  & 3849 & Gennaro \\  
         F150W & JWST & NIRCam  &  20 x 945 s & July 6, 2024  & 3849 & Gennaro\\ 
        \hline
    \end{tabular}
    \label{tab3}
\end{table*}

To investigate the binaries and low-mass stars of Bo\"otes\,I, we used F606W images from the Wide Field Channel of the Advanced Camera for Surveys \citep[ACS/WFC;][]{ford2003} on board the HST, together with F322W2 images from the Near-Infrared Camera \citep[NIRcam;][]{rieke2023} on board the JWST. As discussed in detail in Sects.~\ref{sec_met} and \ref{sec_bin}, the CMD based on these filters maximizes the separation between single and binary stars and is highly sensitive to the chemical composition of M dwarfs. In addition, we used data collected with the ACS/WFC F814W filter and the NIRCam F150W filter to derive stellar proper motions and to separate Bo\"otes\,I members from field stars.

\begin{figure*}
    \centering
    \includegraphics[width=1.\linewidth]{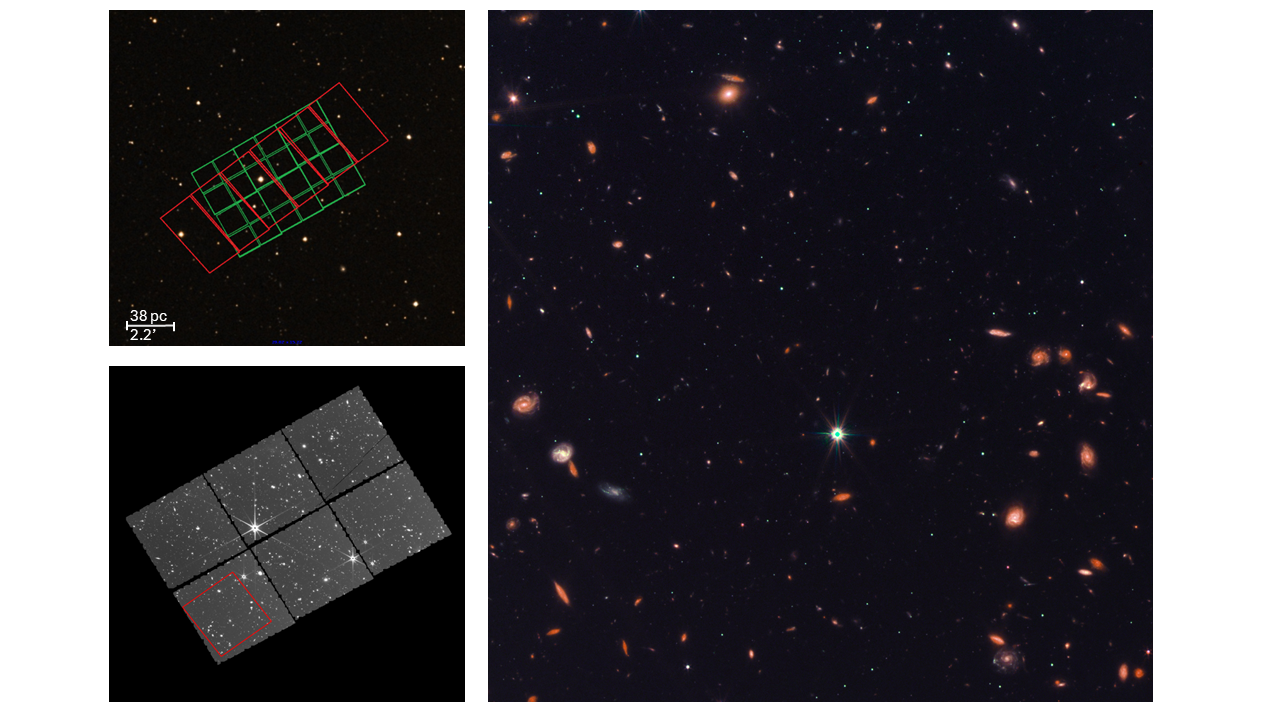}
    \caption{
Footprints of the HST and JWST images used in this work are shown in red and green, respectively, in the top-left panel. The bottom-left panel displays the stacked NIRCam/F322W2 image used in our analysis. The right panel shows a three-colour composite image of the region marked with a red square in the bottom-left panel, where the blue, green, and red channels correspond to the stacked F606W, F814W, and F322W2 images, respectively.
}

    \label{footprint}
\end{figure*}

Because our analysis focuses on precise photometry of faint stars, we considered only the three fields observed in F606W during both GO programs~12549 and~15317. Figure~\ref{footprint} illustrates the footprints of the images used for photometry, and Table~\ref{tab3} summarizes the main properties of the full dataset.

Stellar photometry and astrometry was carried out using Jay Anderson's KS2 software package, an advanced evolution of the ACS/WFC reduction code presented by \citet{anderson2008}. KS2 processes all exposures simultaneously and implements three complementary photometric strategies, each optimized for a different stellar brightness regime \citep[e.g.\,][]{sabbi2016a, bellini2017a, milone2023}.

\textit{Method I} detects sources that produce a significant peak within a 5$\times$5 pixel region after neighbor subtraction. For each star, KS2 measures flux and position in every exposure by fitting an effective point-spread function (ePSF; \citealt{anderson2000}) tailored to the star’s detector location. The sky background is estimated in an annulus spanning 4 to 8 pixels, and final measurements are obtained by averaging results from all images.

\textit{Methods II and III} are designed for stars too faint for robust PSF fitting. Both begin by subtracting neighboring stars and then derive fluxes from aperture photometry.
\textit{Method II} performs weighted aperture photometry in a 5$\times$5 pixel box, reducing the contribution of pixels affected by crowding and adopting the same sky estimate as Method I.
\textit{Method III} is optimized for very crowded regions and for faint stars with numerous exposures available. It uses a circular aperture of radius 0.75 pixels and determines the local background from an annulus between 2 and 4 pixels from the stellar position.
KS2 also offers a suite of diagnostics to assess photometric quality. To ensure high-precision results, we retained only isolated stars that are well fitted by the ePSF model (see Sect. 2.4 of \citealt{milone2023}).

 The photometry was calibrated to the Vega magnitude system following the procedure described in \cite{milone2023}, using the appropriate encircled-energy corrections and zero points provided by STScI.\footnote{STScI calibration resources are available at:\\
ACS:\,\,\url{www.stsci.edu/hst/instrumentation/acs/data-analysis/photometric-calibration}; NIRCam: \url{https://jwst-docs.stsci.edu/jwst-near-infrared-camera/nircam-calibrations}.} We accounted for pixel--area variations and corrected the coordinates for geometric distortion using the solutions of \citet{anderson2022a} for ACS/WFC, and those provided by Jay Anderson for NIRCam data.\footnote{\url{https://www.stsci.edu/stsci-research/research-directory/jay-anderson}}. Finally, 
to derive the proper motions of stars in the field of view of Bo\"otes\,I, we followed the procedure by \cite{milone2023}. In a nutshell, we first derived proper motions relative to the UFD by comparing the positions of stars at different epochs, integrating photometry from HST and JWST. By analyzing these proper motions, we distinguished the majority of Bo\"otes\,I members from field stars. Stars with proper motions deviating by more than three times the galaxy's proper-motion dispersion were excluded from the analysis presented in this paper \citep[see][for details]{milone2023}.

We conducted artificial-star (AS) tests to estimate photometric errors and to generate the simulated CMD. Following the method outlined by \citet{anderson2008}, we created a list of $10^{6}$ ASs. These ASs were designed to mimic the radial distribution and luminosity function of the observed stars, and were placed along the fiducial line from the base of the RGB to the bottom of the MS.

To derive magnitudes and positions for the ASs, we employed the KS2 program, using the same procedures applied to the real stars. Our analysis focused exclusively on relatively isolated ASs that exhibited good PSF fits and satisfied the same selection criteria adopted for the real stars.

\begin{figure}
    \centering
    \includegraphics[width=1.\linewidth]{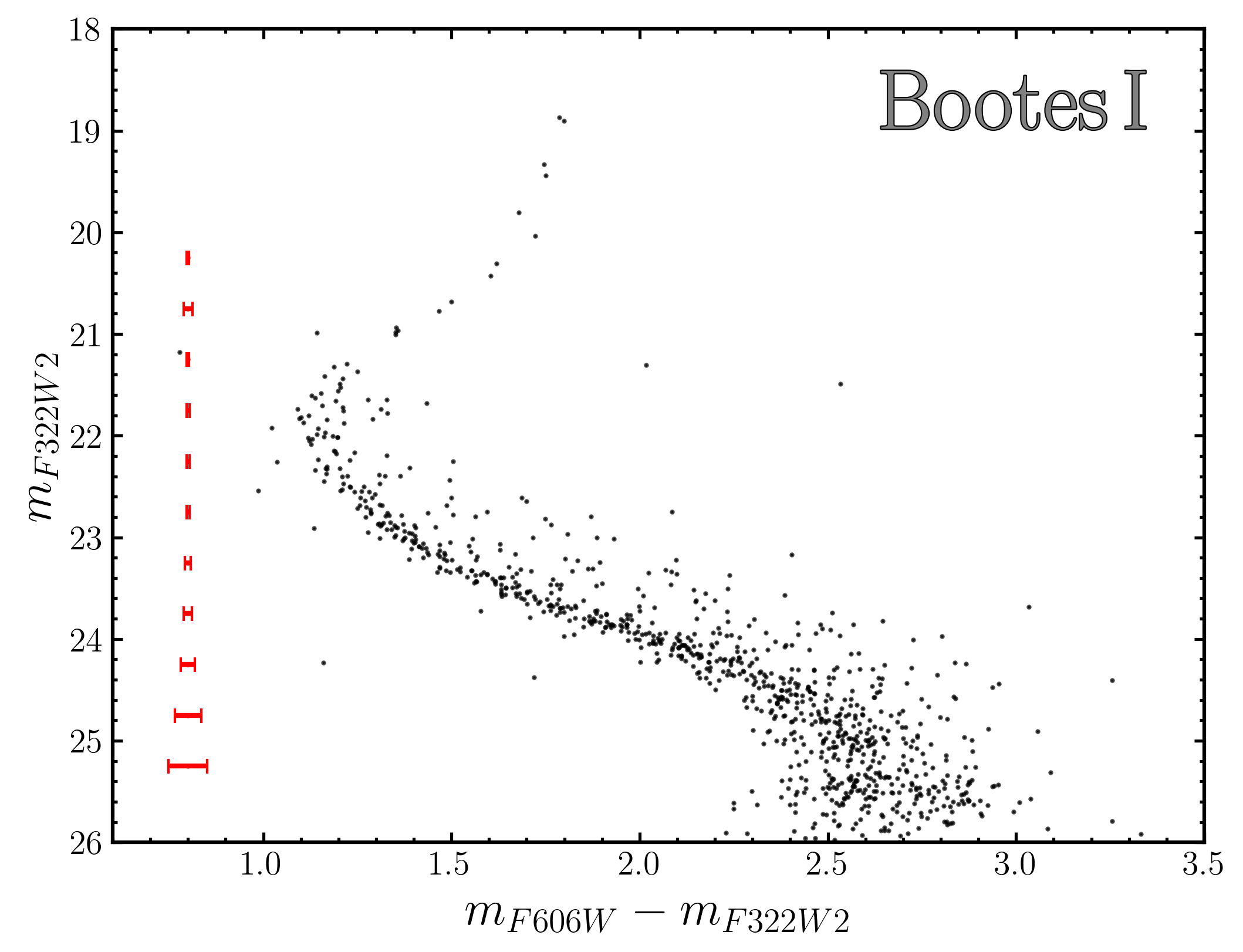}
    \caption{$m_{\rm F322W2}$ vs.\,$m_{\rm F606W}-m_{\rm F322W2}$ CMD of Bo\"otes I. The red bars indicate the average photometric errors.}
    \label{cmd0}
\end{figure}

\section{The color-magnitude diagram of Bo\"otes\,I}
\label{sec_cmd}
Figure~\ref{cmd0} shows the $m_{\mathrm{F322W2}}$ versus $m_{\mathrm{F606W}}-m_{\mathrm{F322W2}}$ CMD of Bo\"otes~I, together with the average error bars, derived from ASs. The CMD is predominantly populated by MS stars, which define a well-visible sequence extending for approximately four F322W2 magnitudes below the turn-off ($m_{\mathrm{F322W2}} = 25.5$~mag). A significant population of MS--MS binary systems is clearly visible on the red side of the MS, while the sub-giant branch (SGB) and RGB are sparsely populated in the upper part of the diagram. There is no evidence for a large magnitude spread among SGB stars.

The MS color broadening varies noticeably with magnitude: it is approximately 0.03~mag around the turn-off, increases to 0.08~mag at $m_{\mathrm{F322W2}} \sim 23.3$~mag, where it becomes comparable to that expected from observational uncertainties alone, and then suddenly increases for $m_{\mathrm{F322W2}} \gtrsim 24.0$~mag.

Section ~\ref{sec_met} exploits the faint MS stars to constrain the metallicity distribution of stars in Bo\"otes~I, taking advantage of the fact that their color is strongly correlated with the chemical composition of this UFD.
In Sect.~\ref{sec_bin}, we take advantage of the narrow and well-defined upper MS to estimate the binary fraction.

\section{Metallicity distribution}
\label{sec_met}
\begin{figure*}
    \centering
    \includegraphics[width=1\linewidth]{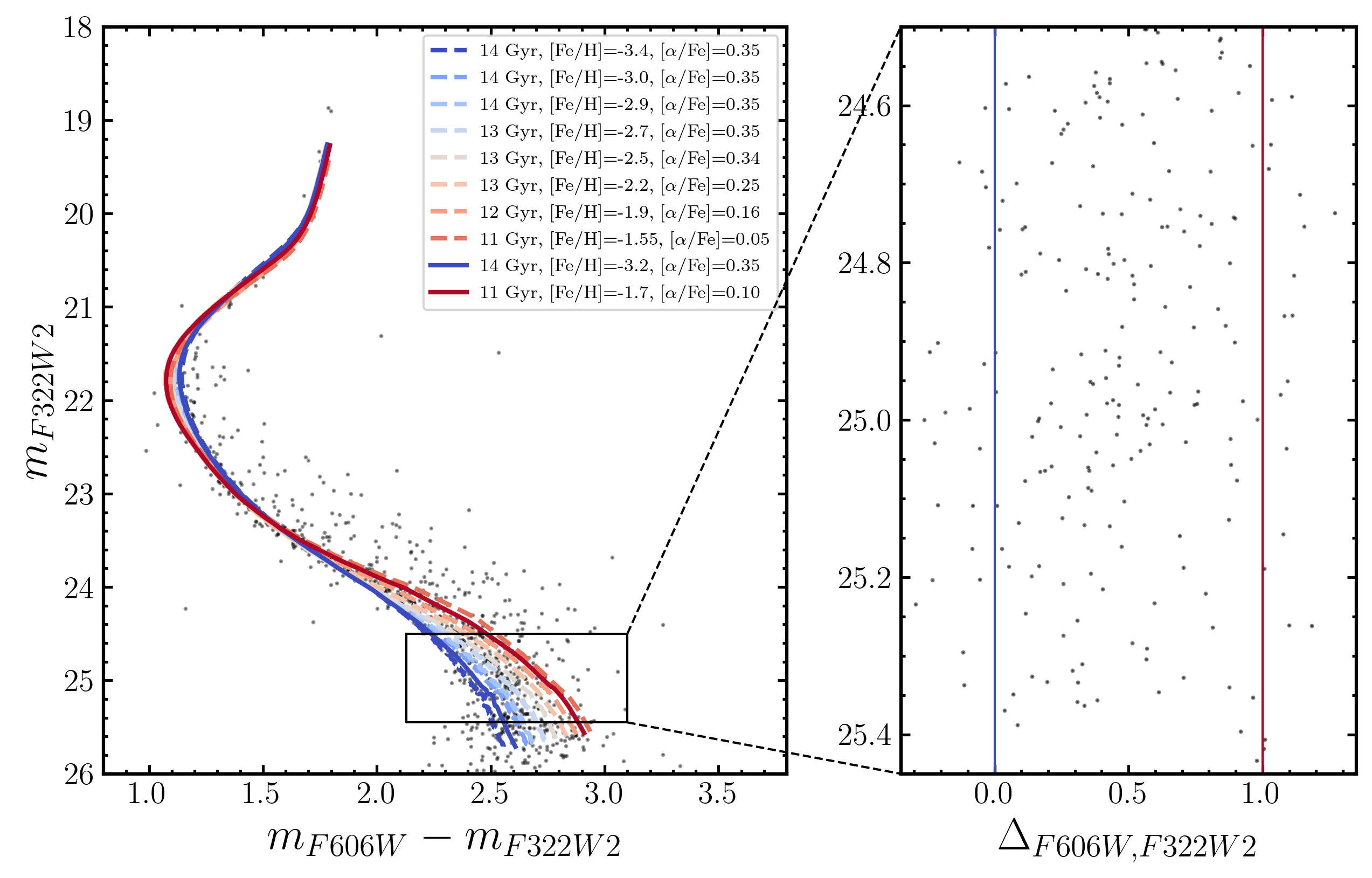}
\caption{
$m_{\rm F322W2}$ vs.\,$m_{\rm F606W}-m_{\rm F322W2}$ CMD of Bo\"otes\,I (left panel) and $m_{\rm F322W2}$ vs.\,$\Delta_{\rm F606W,F322W2}$ verticalized diagram of the lower MS (right panel). We superimpose on the CMD nine isochrones with varying [Fe/H] and [$\alpha$/Fe], as indicated in the inset. The isochrones used to verticalize the diagram are shown with continuous lines.
}

    \label{cmd}
\end{figure*}

To better understand the CMD of Bo\"otes~I, we compare in the left panel of Fig.~\ref{cmd} the observed CMD with isochrones. The adopted metallicity values range from [Fe/H]\,=\,$-3.4$ to $-1.55$\footnote{In selecting the metallicity interval for the isochrone, the lower limit was set by the unavailability of isochrones at very low metallicities, while the onset of the binary region constrained the upper limit. To exclude most binary stars, we therefore restricted our analysis to metallicities below -1.55.}, whereas the corresponding values of [$\alpha$/Fe] for each isochrone are derived from the relation  between [$\alpha$/Fe] and [Fe/H] inferred from \citet[][see their Fig.~8]{frebel2016}:

\begin{gather}
\label{eq1}
\rm[\alpha/Fe]=
\begin{cases}
     0.35 \quad  \hspace{2.5cm}\rm [Fe/H] \leq -2.5 \\
   -0.31 \cdot \rm[Fe/H] - 0.43 \quad \rm[Fe/H] > -2.5 \\
\end{cases}
\end{gather}
The isochrones with [Fe/H]~$\geq -3.2$ are taken from the BaSTI database \citep{hidalgo2018a}\footnote{http://basti-iac.oa-abruzzo.inaf.it/isocs.html}, while those for more metal-poor populations were derived by linearly interpolating the colors and magnitudes between the [Fe/H]~=~$-$3.0 and $-$3.2 BaSTI isochrones. 

We obtained the best fit by using a distance modulus of $(m-M)_{0}=18.86$, a foreground reddening of $E(B-V)=0.10$~mag, and ages of 14--11~Gyr (see Appendix A. for more details on ages selection).

Notably, these isochrones encompass the bulk of stars along the entire MS and indicate that the observed variation in MS width as a function of magnitude is consistent with stellar populations characterized by different metallicities. In particular, the isochrones indicate that the color spread of the lower MS ($m_{\mathrm{F322W2}} \gtrsim 24.4$) is strongly driven by the chemical composition.
Conversely, the color of the MS portion between $m_{\mathrm{F322W2}} \sim 22.5$ and 24.4 is much less sensitive to metallicity changes, since the isochrones shown in Fig.~\ref{cmd} exhibit a reduced color separation and even overlap within this magnitude range.

To investigate the metallicity distribution of Bo\"otes\,I stars, we selected MS stars within the CMD region defined by $24.5 < m_{\mathrm{F322W2}} < 25.5$~mag, where stellar colors are strongly affected by metallicity.  
Following the procedure described in \citet{milone2017}, we verticalized the selected stars such that the isochrones with [Fe/H]~$=-3.2$ and [Fe/H]~$=-1.7$ (blue and red solid lines in Fig.~\ref{cmd}) are mapped onto vertical lines with abscissae equal to 0 and 1, respectively.  
For each star, we computed the pseudo-color parameter:
\begin{equation}
    \Delta_{\mathrm{F606W,F322W2}} = 
    \frac{X - X_{\mathrm{red\,isochrone}}}
         {X_{\mathrm{red\,isochrone}} - X_{\mathrm{blue\,isochrone}}},
\end{equation}
where $X$ represents the color of the star in the CMD.

\begin{figure}
    \centering
    \includegraphics[width=0.8\linewidth]{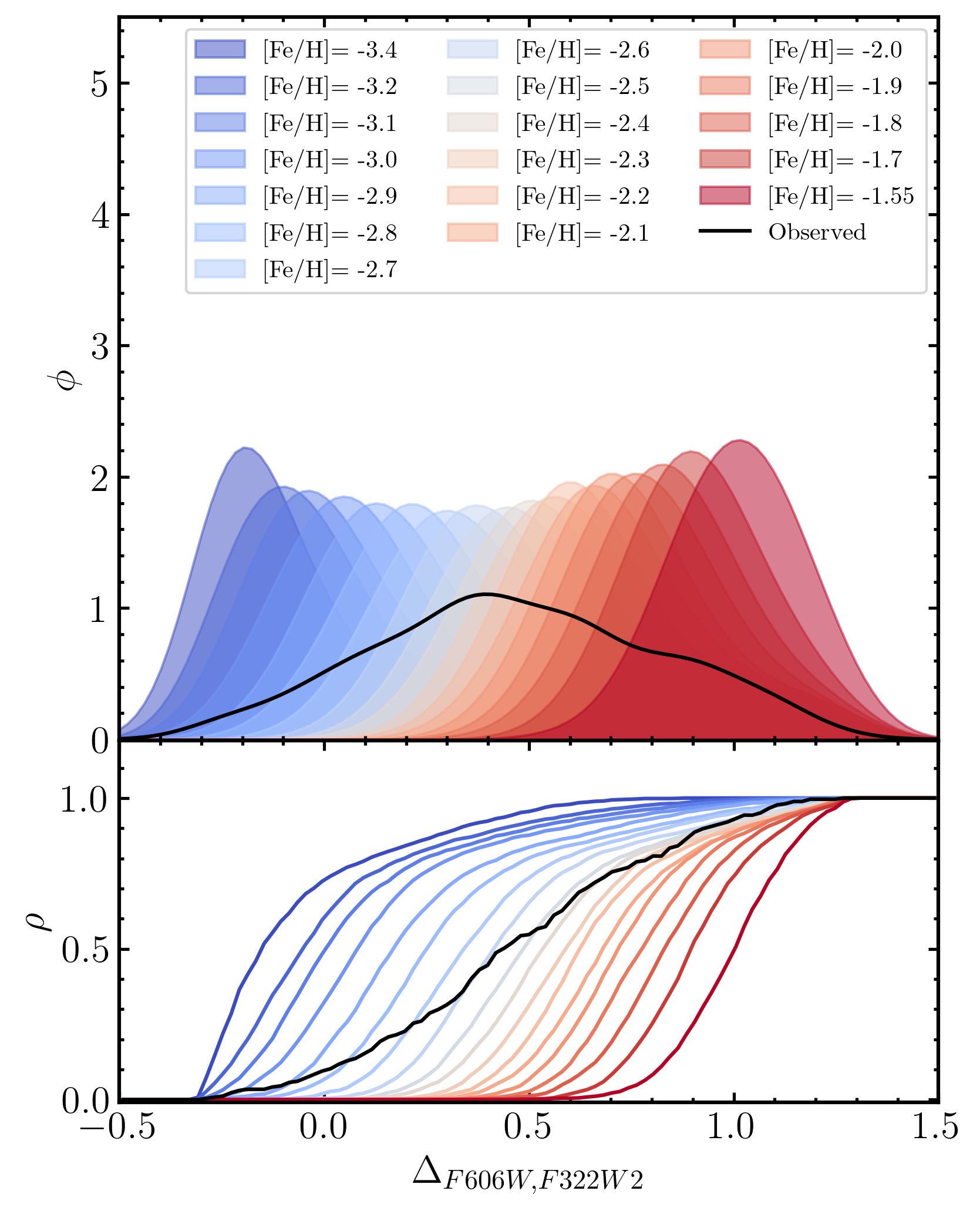}
    \caption{
Kernel (top) and cumulative (bottom) distributions of $\Delta_{\rm F606W,F322W2}$ for faint MS stars. The black curves are derived from the observed Bo\"otes\,I members, while the colored curves are obtained from the simulated CMDs.
}
    \label{kde_cdf}
\end{figure}

The resulting verticalized diagram is shown in the right panel of Fig.~\ref{cmd}. 
The parameter $\Delta_{F606W,F322W2}$ serves as a proxy for the metallicity distribution. 
Figure~\ref{kde_cdf} displays the kernel ($\phi$), evaluated using a Gaussian kernel with a bandwidth of 0.1, and the cumulative ($\rho$) distributions of $\Delta_{F606W,F322W2}$ as solid black lines.
These distributions are approximately symmetric, with the majority of MS stars exhibiting intermediate values of $\Delta_{F606W,F322W2}$.

To gather information on metallicity, we compared the observed photometry with simulated diagrams.
We used ASs to generate seventeen simulated CMDs corresponding to simple stellar populations with metallicities ranging from [Fe/H] = $-$3.2 to $-$1.7 in steps of 0.1~dex, assuming the relation between [$\alpha$/Fe] and [Fe/H] (Eq. \ref{eq1}) inferred from \cite{frebel2016}. In addition, we produced two limiting simulated CMD corresponding to stellar populations with [Fe/H] = $-$3.4 and $-$1.55.
A binary fraction of 0.30 was adopted, with a flat mass--ratio distribution. 
The input colors and magnitudes of the ASs were obtained from the BaSTI isochrones.
We derived the $\Delta_{\rm F606W,F322W2}$ pseudo-color for stars in each simulated diagram by applying the same methodology and adopting the same magnitude range as for the observed stars. In Fig.\,\ref{kde_cdf}, we compare the kernel-density and cumulative distributions of the observed stars (black lines) with those derived from the simulated CMDs.  

To estimate the metallicity distribution of Bo\"otes\,I, we adapted to the sample of MS stars identified in Fig.\,\ref{cmd} the method introduced by \citet{stauffer1980} and widely adopted for studying stellar populations \citep[e.g.][]{cignoni2010a, cordoni2022a, legnardi2022}. In essence, we compared the cumulative $\Delta_{F606W,F322W2}$ distribution of large numbers of simulated stellar populations with the observed one.

We constructed a composite simulated population by summing the individual model populations, assigning to each population \( j \) a weighting factor \( c_{j} \) that defines its fractional contribution to the total number of stars in the resulting $\Delta_{F606W,F322W2}$ cumulative distribution. The coefficient \( c_{j} \) ranges between 0 and 1. The composite cumulative distribution was then quantitatively compared with the observed distribution through a classical \(\chi^{2}\) minimization procedure. This optimization was performed using the open-source \texttt{pyGAD} package\footnote{We used \texttt{pyGAD} (\url{https://pygad.readthedocs.io/}), a genetic algorithm Python library that reduces the likelihood of convergence to a local rather than a global minimum. The coefficients were normalized using a softmax function, ensuring that their sum equals unity and can be interpreted as fractional contributions resulting from the minimization.}, which provides the array of coefficients \( c_{j} \)  describing the relative contribution of each model population to the best-fitting cumulative distribution of the Bo\"otes\,I stars.

\begin{figure*}
    \centering
    \includegraphics[width=1.\linewidth]{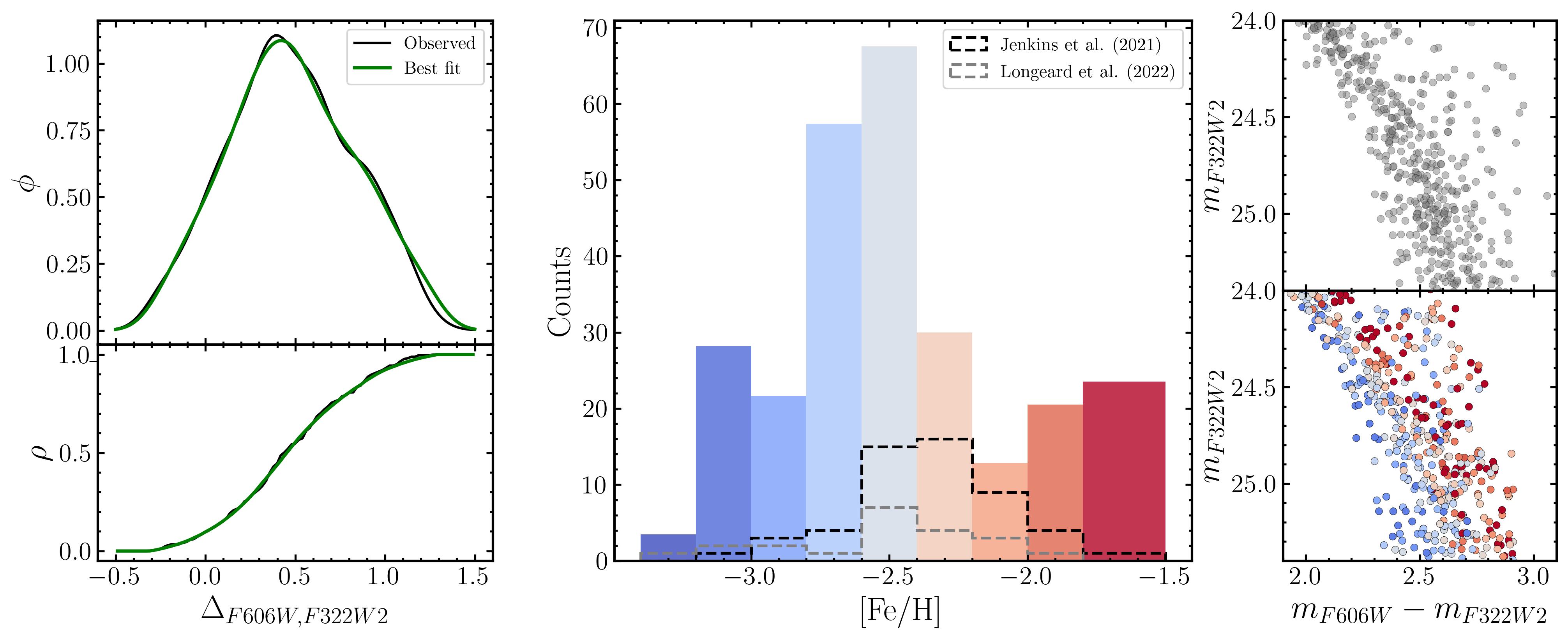}
    \caption{
\textit{Left:} Comparison between the observed (black) and best-fit simulated (green) kernel (top) and cumulative (bottom) distributions of $\Delta_{\rm F606W,F322W2}$. 
\textit{Middle:} [Fe/H] histogram that best reproduces the observations of Bo\"otes\,I. The dashed black and gray distributions are from a spectroscopic survey from \cite{jenkins2021} and \cite{longeard2022}, respectively.
\textit{Right:} Comparison between the observed (top) and simulated (bottom) CMDs, zoomed in on the MS region used to infer the metallicity distribution. 
The colors of the simulated stars indicate their iron abundance, as shown in the middle panel.
}

    \label{met}
\end{figure*}

The bottom-left panel of Fig.~\ref{met} compares the observed cumulative distribution with the best-fit model. For completeness, the upper-left panel of the same figure shows the comparison between the observed kernel-density distribution and that derived from the best-fit simulation.  
As shown in Table\,\ref{tab2}, the best-fitting simulated distribution gives a mean iron abundance of [Fe/H]~$=-2.53\pm0.05$ and a dispersion of $0.41\pm0.03$~dex. These values are consistent with spectroscopic measurements, 
[Fe/H]~$=-2.60 \pm 0.03$~dex, with a metallicity dispersion of $0.34\pm0.03$~dex \citep{longeard2022}.

The corresponding metallicity histogram is shown in the middle panel of Fig.~\ref{met}, revealing a main peak near the average metallicity value.  
The most metal-poor stars have [Fe/H]~$\sim-3.4$, and their number gradually increases up to the maximum around the mean metallicity.  
The number of stars then decreases toward higher metallicities, with a tail of stars having [Fe/H]~$\gtrsim-2$. In particular, we found that the majority of the stars $\sim$85\% have [Fe/H] $< -2$, and that roughly $\sim$17\% have [Fe/H] $< -3$.
Finally, the right panels of Fig.~\ref{met} compare the observed (top) and simulated (bottom) CMDs, zoomed in on the faint end of the MS, highlighting the strong similarity between the two diagrams. \\ To assess the impact of the adopted $[\alpha/\mathrm{Fe}]$ distribution, we re-derived stellar metallicities assuming two extreme and constant values, $[\alpha/\mathrm{Fe}]=0.35$ and $[\alpha/\mathrm{Fe}]=0.00$~dex. Although neither of these simplified prescriptions accurately reproduces the observed $[\alpha/\mathrm{Fe}]$ distribution in Bo\"otes~I, they provide a useful way to quantify the sensitivity of our results to this assumption.
For $[\alpha/\mathrm{Fe}]=0.35$, the inferred metallicity distribution closely resembles the reference distribution derived in this work, but is shifted toward slightly lower metallicities, with a mean $[\mathrm{Fe}/\mathrm{H}]=-2.60\pm 0.02$ and a dispersion of $\sim 0.3$~dex.
In contrast, adopting a solar-scaled composition ($[\alpha/\mathrm{Fe}]=0.00$) leads to a substantial shift of the metallicity distribution toward higher values, yielding a mean $[\mathrm{Fe}/\mathrm{H}]=-2.10\pm 0.02$ and a comparable dispersion of $\sim 0.3$~dex. This result is in clear disagreement with both our fiducial metallicity distribution and those obtained from spectroscopic studies \citep{norris2010b, koposov2011, gilmore2013, ishigaki2014, jenkins2021, longeard2022, sandford2025}. Such a discrepancy is expected, since assuming solar-scaled abundances for all stars is inconsistent with the enhanced $[\alpha/\mathrm{Fe}]$ ratios observed in Bo\"otes~I.

\section{The Binary Map}
\label{sec_bin}

To investigate the population of binaries along the MS of Bo\"otes\,I, we introduce the \textit{Binary Map}, a new diagnostic tool analogous to the chromosome map \footnote{The chromosome map \citep{milone2015,milone2017} is a powerful diagnostic tool for studying stellar populations in GCs. It is a pseudo–two-color diagram that is highly sensitive to the chemical differences between distinct stellar populations.} of GCs, designed to distinguish binaries with different mass ratios.

This diagram is based on the well-known photometric properties of unresolved binary systems composed of two MS stars, which appear as single point-like sources with a total magnitude

\begin{equation}
    m_{\mathrm{bin}} = m_{1} - 2.5 \log \left( 1 + \frac{F_{2}}{F_{1}} \right),
\end{equation}

where $m_{1}$ is the magnitude of the primary (brightest) star, and $F_{1}$ and $F_{2}$ are the fluxes of the primary and secondary components, respectively. 
In a simple stellar population, the flux of MS stars depends on their mass, following a specific mass-luminosity relation. As a consequence, the total magnitude of a binary system composed of two MS stars can be uniquely determined by the mass of the primary component and the mass ratio, $q = M_{2} / M_{1}$, where $M_{1}$ and $M_{2}$ are the masses of the primary and secondary stars.

\begin{figure}
    \centering
    \includegraphics[width=1.\linewidth]{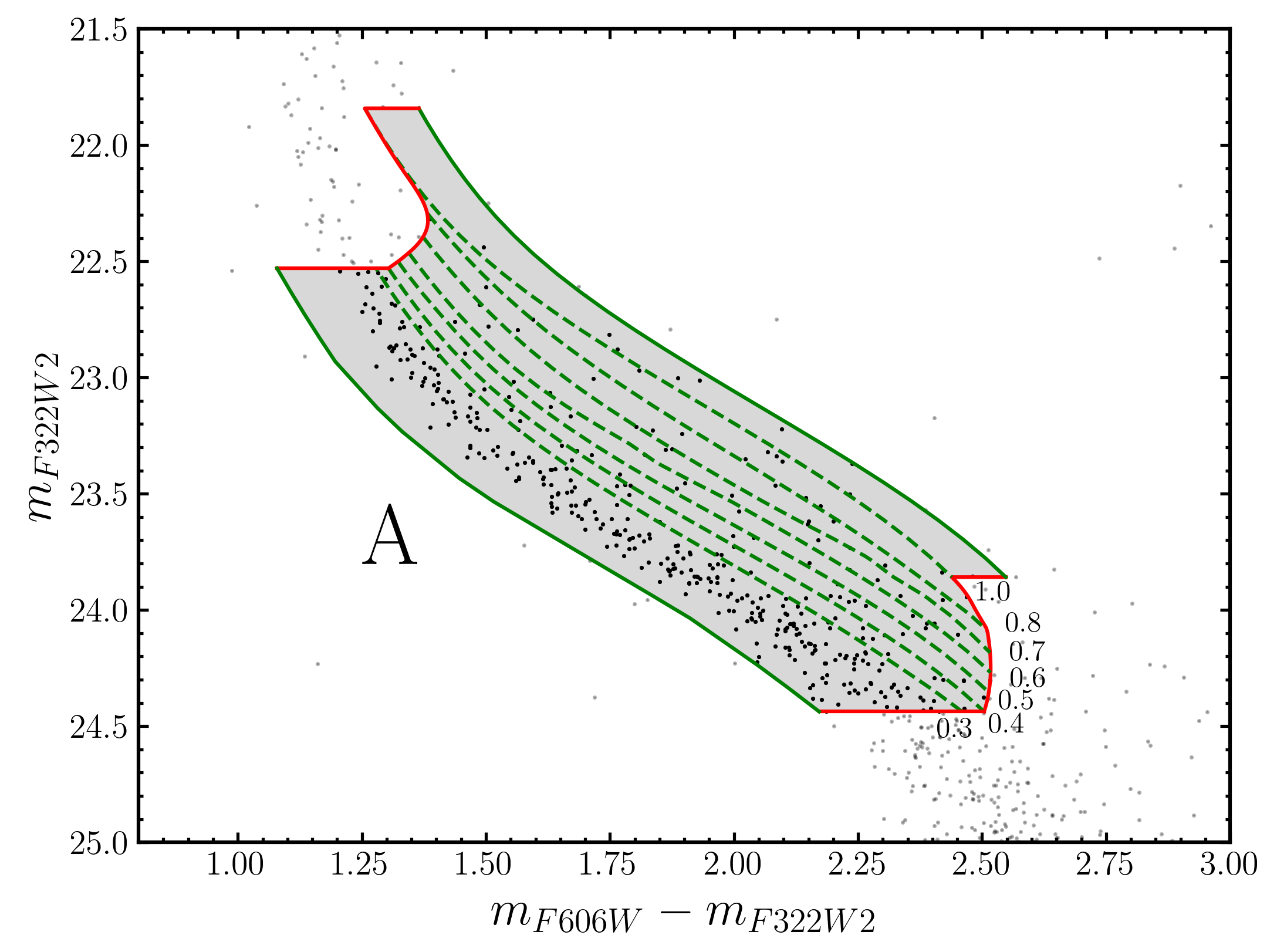}
\caption{
$m_{\rm F322W2}$ vs.\,$m_{\rm F606W}-m_{\rm F322W2}$ CMD of Bo\"otes\,I, where the grey shaded area indicates the region used to study binaries and derive the \textit{Binary Map}. The nine green lines delimit the eight subregions (R1--R8) adopted for the binary analysis. The bluest line represents the fiducial sequence of single stars, shifted to the blue to encompass the bulk of MS stars, while the reddest line corresponds to the fiducial of equal-mass binaries, shifted to the red to include most binaries. The remaining lines represent fiducial sequences of binaries with mass ratios ranging from 0.3 to 1.0, as indicated.
}

    \label{cmd_strips}
\end{figure}

To construct the \textit{Binary Map}, we used the setup illustrated in Fig.\,\ref{cmd_strips}, where the shaded area superimposed on the CMD of Bo\"otes\,I highlights the region, hereafter referred to as region~A, used to study the binary systems. 
This region includes single stars with $22.5 < m_{\rm F322W2} < 24.5$ and binary systems whose primary component falls within the same magnitude interval. 
We selected this portion of the CMD because of the partial overlap between isochrones of different metallicities (Fig. \ref{cmd}), so that binaries are well separated from single stars.

As shown in Fig.\,\ref{cmd_strips}, we defined nine green reference lines, hereafter referred to as lines~1--9. 
Line~1, which represents the left boundary of region~A, corresponds to the fiducial line of Bo\"otes\,I shifted toward the blue by three times the average color error ($\sim$0.16\,mag) in order to enclose the bulk of MS stars. 
Lines~2 to~8, the dashed ones, correspond to the fiducial sequences of binary systems with mass ratios $q = 0.3$, 0.4, 0.5, 0.6, 0.7, 0.8, and 1.0, respectively.

These fiducial lines were derived following the same procedure adopted in our previous works \citep[e.g.][]{milone2012,muratore2024}, using the MS fiducial line and the mass--luminosity relation provided by the best-fit isochrone to compute the colors and magnitudes corresponding to binaries with different $q$ values. 
Finally, line~9 represents the fiducial sequence of equal-mass binaries shifted toward the red by two times the average color error ($\sim$0.11\,mag), to include binaries with large mass ratios that appear redder due to observational uncertainties.
These nine lines define eight regions, namely $R_1$--$R_8$. Region $R_1$ includes the bulk of single stars and binaries with a mass ratio smaller than 0.3. 
Regions~$R_2$--$R_7$ contain binaries in successive $0.1$-wide $q$ bins from $0.3$ to $0.8$, while region~$R_8$ covers $0.8\leq q\leq1.0$ (width $0.2$).
Region 8 mostly hosts binaries that are shifted to the red by photometric uncertainties. 

We applied a two-step verticalization of the diagram following the procedure used to derive the chromosome maps of GCs. 
In the first step, we normalized the colors such that lines~1--9 are transformed into vertical lines. 
To achieve this, we adapted Eq.~1 from \citet{milone2025a} to each region ($R_{i}$, $i=1$--8) and derived the quantities

\begin{gather}
\delta_{F606W,F322W2}^{i} =
\begin{cases}
     W_{i} \, \frac{X - X_{\mathrm{line}\,i}}{X_{\mathrm{line}\,i} - X_{\mathrm{line}\,i+1}} + \sum_{j=1}^{i-1} W_{j} \quad \text{if } X \in R_i \\
   \hspace{2cm}  0  \hspace{1.5cm} \quad \text{else }
\end{cases}
\end{gather}

where $X = m_{\mathrm{F606W}} - m_{\mathrm{F322W2}}$, and $W_{i}$ is the color separation between lines~$i+1$ and~$i$, measured at $m_{F322W2}=$23.5\,mag. 
We then combined these quantities to define the pseudo-color:

\begin{equation}
    \Delta_{F606W,F322W2}^{Bin} = \sum_{i=1}^{8} \delta_{F606W,F322W2}^{i}
\end{equation}

In the second step, the verticalization is performed in magnitude. Specifically, for each region, we define the quantity
\begin{equation}
\label{eq3}
    \Delta_{F322W2} = W_{y} \, \frac{Y - Y_{\mathrm{top\,line}, i}}{Y_{\mathrm{top\,line}, i} - Y_{\mathrm{bottom\,line}, i}},
\end{equation}
where $Y = m_{\mathrm{F322W2}}$, and $W_{y}$ denotes the magnitude interval of region~$R_1$. The top line corresponds to red top line from $m_{\mathrm{F322W2}} = 22.5$ in region~$R_1$, to the segment with $m_{\mathrm{F322W2}} = 22.5 - 0.75$ in region~$R_8$, tracing binaries whose primary star has $m_{\mathrm{F322W2}} = 22.5$ and mass ratios between 0.3 and~1.0 in the remaining regions.  
The bottom red line is defined analogously, but for $m_{\mathrm{F322W2}} = 24.4$.

The resulting $\Delta_{F322W2}$ versus $\Delta_{F606W,F322W2}^{Bin}$ \textit{Binary map} is shown in the left panel of Fig.\,\ref{binary_map}, together with the $\Delta_{F606W,F322W2}$ histogram. The majority of stars lie within region~1--2, but a well-populated tail extends into regions~3--8, confirming that Bo\"otes~I hosts a significant fraction of binaries with $q > 0.4$.

\subsection{The fraction of binaries in Bo\"otes\,I}

To reproduce the observed stellar distribution across the \textit{Binary map}, we adopted an iterative procedure based on ASs. The purpose of this method is to generate a simulated sample whose recovered stars reproduce, region by region, the number of observed stars, $N_i$.

In the first iteration, we generated for each region a number of input ASs equal to half of the observed count, that is,
\[
n_{\mathrm{INP},i,1} = \frac{N_i}{2}.
\]
Each AS was assigned input magnitudes and colors that placed it within the region boundaries. Due to observational uncertainties and photometric scatter, some ASs injected within a given region were later recovered in different regions of the \textit{Binary map}. The number of ASs actually recovered in each bin after this first iteration is denoted by $n_{\mathrm{GEN},i,1}$. 

In the second iteration, we injected an additional set of ASs. The number of newly added stars in each bin was determined as
\[
n_{\mathrm{INP},i,2} = \frac{N_i - n_{\mathrm{GEN},i,1}}{2},
\]
so as to gradually compensate for bins where the number of recovered stars was smaller than the observed value. The cumulative number of injected stars in each bin after this step is therefore
\[
n_{\mathrm{INP},i,\mathrm{TOT}} = n_{\mathrm{INP},i,1} + n_{\mathrm{INP},i,2}.
\]
Once again, we measured the number of recovered ASs in each bin, $n_{\mathrm{OBS},i,2}$, after running the full reduction and selection procedure.

The process was repeated iteratively, changing the number of injected stars in each bin according to the residual difference between the observed and recovered counts. Convergence was reached when, for all bins,
\[
\big|\, n_{\mathrm{OBS},i,X} - N_i \,\big| < 1,
\]
 This iterative approach ensures that the AS sample reproduces the observed distribution of stars across the \textit{Binary map} while accounting for photometric scatter and completeness effects.

\begin{figure*}
    \centering
    \includegraphics[width=1.\linewidth]{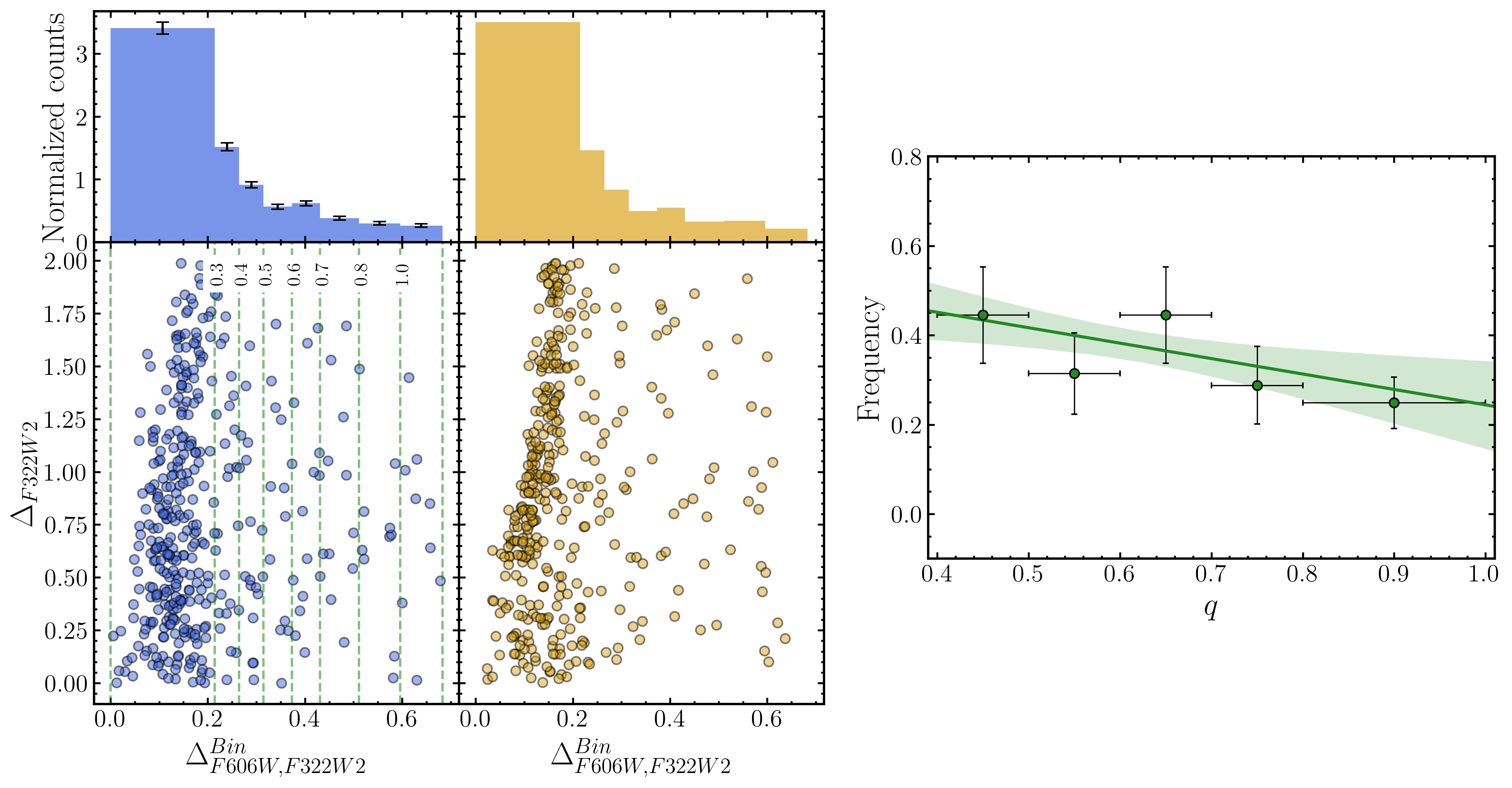}
    \caption{Binary maps and mass-ratio distribution for Boötes I. Left: the observed \textit{Binary map} (in blue), with the corresponding $\Delta_{F606W,F322W2}$ distribution shown in the top panel. The dashed vertical lines indicate the nine reference loci used to construct the map, including the fiducial sequences of binary systems with different mass ratios, $q$, whose values are labelled in the figure. Middle: the best-fit \textit{Binary map} (in gold). Right: the mass-ratio distribution derived from the best fit; the green line and shaded region indicate the linear fit and its associated uncertainty. }
    \label{binary_map}
\end{figure*}

Figure~\ref{binary_map} compares the observed \textit{Binary map} of Bo\"otes~I (left) with the best-fit simulated one (middle). The corresponding $\Delta_{\mathrm{F606W,F322W2}}^{\mathrm{Bin}}$ histogram distributions are shown in the top panels.
This procedure provides the numbers of single stars and binaries with their respective mass ratio. 
 
The right panel of Fig.~\ref{binary_map} illustrates the binary frequency, defined as the fraction of binaries in five mass-ratio intervals divided by the width of each interval, as a function of mass ratio. The best-fit straight line has a slope of $-0.30 \pm 0.15$, which is consistent with a flat distribution at the $2\sigma$ level. We consider only binaries with mass ratios larger than 0.4 because the small color separation makes it challenging to robustly distinguish low-mass-ratio binaries from single stars.
We obtained an almost flat mass-ratio distribution, consistent with the results of \cite{raghavan2010R} and \cite{milone2012} for solar-type stars in the Galactic field and for 29 GCs, respectively.

\begin{figure}
    \centering
    \includegraphics[width=1.\linewidth]{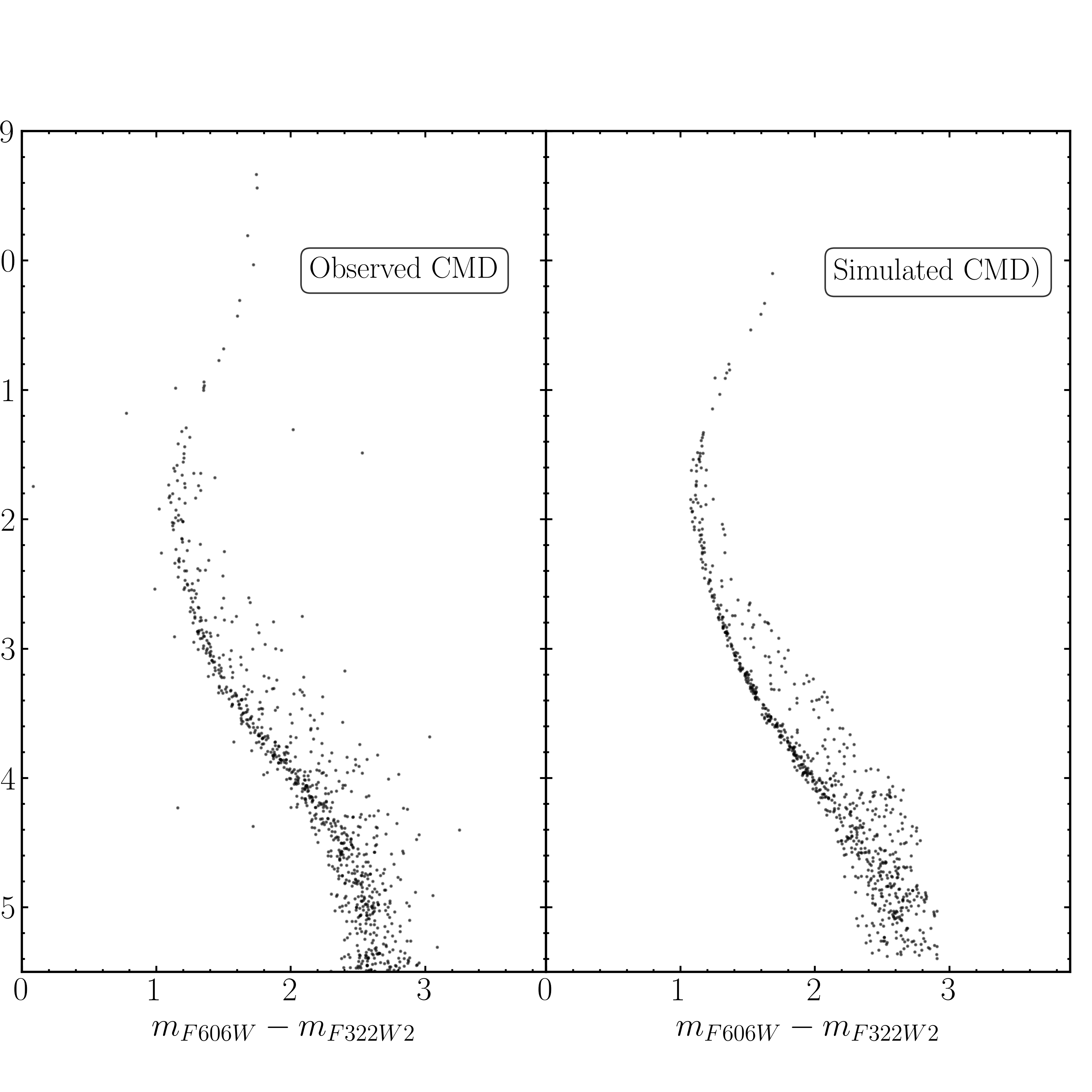}
    \caption{
    Comparison of the observed (left) and simulated (right) CMD of Bo\"otes\,I.
    }
    \label{composite}
\end{figure}

By combining the results from the five bins, we derived a binary fraction with mass ratio larger than 0.4 of $f_{\mathrm{bin}}^{q>0.4} = 0.20 \pm 0.02$. 
Assuming this mass-ratio distribution, we infer an overall binary fraction of $f_{\mathrm{bin}}^{\mathrm{TOT}} = 0.30 \pm 0.03$, taking into account a minimum mass of 0.075 $M_{\odot}$ for the secondary star. 
The derived parameters are summarized in Table~\ref{tab2}.
Figure \ref{composite} shows a comparison between the observed Bo\"otes I and the simulated CMDs with the derived binary fraction, metallicity distribution, and ages from 11 to 14 Gyr.

\subsection{Comparison with  UFDs, clusters and field stars}

\begin{figure*}
    \centering
    \includegraphics[width=1\linewidth]{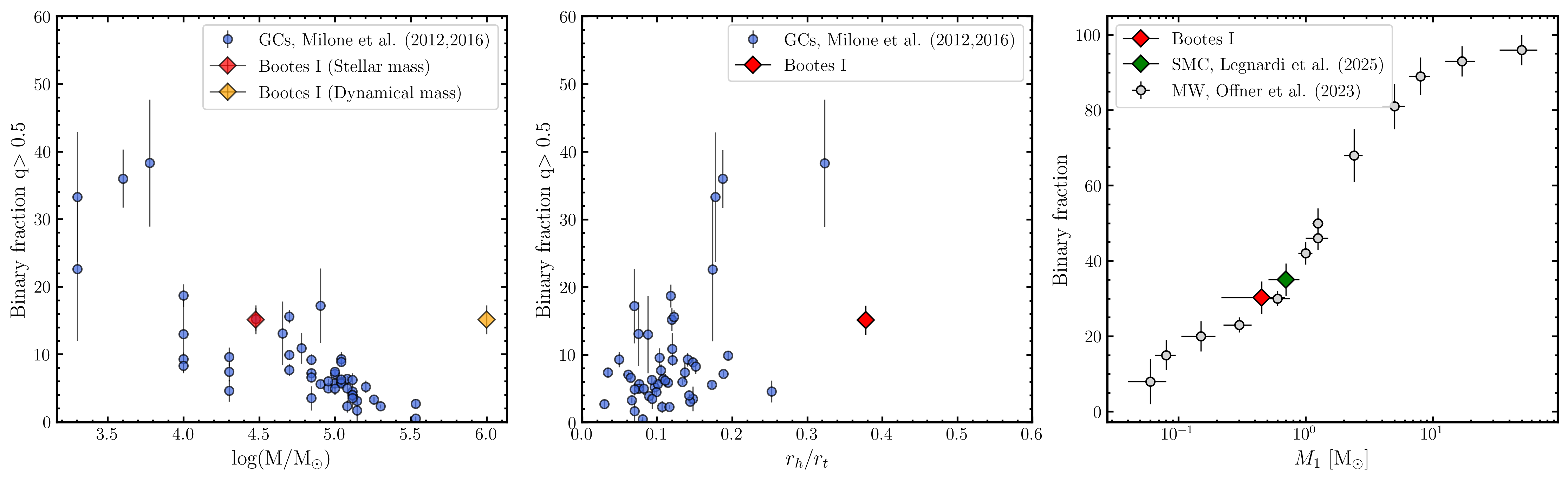}
    \caption{Comparison between Bo\"otes I, Galactic GCs, and field stars. Left: fraction of binaries with mass ratio $q>0.5$ in the cores of GCs \citep{milone2012,milone2016} compared with that in Bo\"otes I. The red and yellow diamonds represent Bo\"otes I, corresponding to the dynamical and stellar masses, respectively, from \cite{martin2008} and \cite{jenkins2021}. GC masses are adopted from \cite{Baumgardt2020}. Middle: binary fraction plotted against the concentration parameter, $r_h/r_t$. Radii are taken from the 2010 edition of the \cite{harris1996a} catalog for GCs, and from \cite{okamoto2012} and \cite{belokurov2006} for Bo\"otes I. Right: total binary fraction as a function of the primary-component mass. Gray dots correspond to Galactic field stars compiled in the review by \cite{offner2023}, and the green diamond indicates the binary fraction measured for the SMC field by \cite{legnardi2025}.}
    \label{fbin_mass}
\end{figure*}

The binary fraction we measure in Bo\"otes\,I is consistent with the spectroscopic estimate of $\sim0.28$ reported by \cite{gennaro2018}. It is also similar to the values found in Leo II and in the inner region of Ursa Minor, which are $\sim0.30$ and $\sim0.33$, respectively \citep{spencer2017, qiu2025}. 
Moreover, the binary fraction of Bo\"otes\,I is comparable to that typically observed in Galactic open clusters \citep{cordoni2023}, and in Magellanic Cloud clusters \citep{mohandasan2024}, and
significantly higher than that found in most GCs \citep{milone2012, milone2016}.  
This comparison is illustrated in the left and central panels of Fig.~\ref{fbin_mass}.  
The left panel shows the anticorrelation between the binary fraction in GCs \citep{milone2012, milone2016} and the mass of the host cluster \citep{Baumgardt2020}.  
Remarkably, when considering only the stellar mass \citep{jenkins2021}, Bo\"otes\,I exhibits a binary fraction similar to that of GCs with comparable masses.  
In contrast, when the dynamical mass \citep{martin2008} is adopted, Bo\"otes\,I deviates from the relation followed by the bulk of GCs.

For completeness, the middle panel of Fig.~\ref{fbin_mass} shows the relation between the binary fraction and the concentration parameter, defined as the ratio between the half-light radius and the tidal radius of the stellar system.  
The concentration parameter is often used to distinguish GCs from UFDs.  
We adopted GC structural parameters from the 2010 edition of the \citet{harris1996a} catalog, and those for Bo\"otes\,I from \citet{okamoto2012} and \citet{koposov2011}.  
As expected, Bo\"otes\,I exhibits a larger $r_{\rm h}/r_{\rm t}$ ratio than that typically observed in GCs.  
There is no clear evidence for a strong correlation between the concentration and the binary fraction, although stellar systems (both Bo\"otes\,I and GCs) with $r_{\rm h}/r_{\rm t} \gtrsim 0.15$ tend to have, on average, higher binary fractions than the remaining GCs.  
Nevertheless, the large scatter in binary fraction among systems with similar $r_{\rm h}/r_{\rm t}$ values prevents us from drawing firm conclusions about any relation between concentration and binary fraction.  

Finally, as shown in the right panel of Fig.~\ref{fbin_mass}, where we plot the binary fraction against the mass of the primary stars, Bo\"otes\,I follows the same trend defined by Milky Way and Small Magellanic Cloud field stars \citep{offner2023, legnardi2025}.

\begin{table}[h]
    \centering
        \caption{Binary fraction with mass ratio larger than 0.4, total binary fraction, slope of the mass ratio distribution, average iron abundance and dispersion of Bo\"otes I. 
        }

    \setlength{\tabcolsep}{8pt} 
    \renewcommand{\arraystretch}{1.3} 
    \begin{tabular}{l l }
        \hline
        Parameter & Value \\
        \hline
        $\langle \rm[Fe/H]\rangle$ & $-2.53 \pm 0.05$   \\
        $\sigma_{\rm[Fe/H]}$ &$ 0.41 \pm 0.03$ \\
        $f_{bin}^{q>0.4}$ & $0.20\pm 0.02$ \\
        $f_{bin}^{\mathrm{TOT}}$ & $0.30\pm 0.03$ \\
        q slope & $-0.30 \pm 0.15$ \\

        \hline
    \end{tabular}
    \label{tab2}
\end{table}

\section{Summary and discussion}
\label{sec_con}
We derived high-precision  HST and  JWST photometry for stars in the core of the UFD galaxy Bo\"otes I, using the F606W band of ACS/WFC and the F322W2 band of NIRCam. The resulting exquisite CMD reveals a remarkably narrow and bright MS, where the observed color spread is comparable with that expected from photometric uncertainties alone over the F322W2 magnitude range $\sim$22.5–24.5. At fainter magnitudes, however, the MS color broadening increases sharply, reaching $\sim$0.4 mag around $m_{\rm F322W2}\sim25.0$. A significant color broadening is also evident near the MS turnoff, which transitions into a narrow and sparsely populated SGB.  Comparison with isochrones indicates that this behavior is consistent with the presence of stellar populations spanning a range of metallicities. Notably, the CMD also displays a prominent binary sequence, formed by pairs of MS stars, located on the red side of the single-star MS. We used this data to infer the metallicity distribution and to investigate the population of binaries.

\begin{itemize}

\item The wide color broadening of stars fainter than $m_{\rm F322W2}\sim24$ provides an opportunity to constrain the chemical composition of stars in Bo{\"o}tes\,I. After accounting for the [$\alpha$/Fe]--[Fe/H] relation inferred from high-resolution spectroscopy \citep{frebel2016}, the color distribution of faint stars allows for an accurate determination of the metallicity distribution \citep[see also][for similar determinations in GCs]{legnardi2022,legnardi2024}. Based on a sample of 265 stars, we infer an average iron abundance of [Fe/H]$=-2.53\pm0.05$, with a dispersion of $0.41\pm 0.03$. The metallicity distribution spans a wide range, from [Fe/H]$\sim-3.4$ up to [Fe/H]$\sim-1.55$, indicating that approximately $\sim$85\% have [Fe/H] $< -2$, and that roughly $\sim$17\% have [Fe/H] $< -3$. 
These findings are consistent and confirm previous high-resolution spectroscopic studies based on smaller samples of RGB stars \citep{jenkins2021, longeard2022, sandford2025}. This approach, applied here for the first time to a UFD, combines photometric and spectroscopic information ($\rm[\alpha/Fe]$-$\rm[Fe/H]$ relation) to analyze a large sample of MS stars. Our analysis showed that this new technique can be used for systems where only a limited number of spectra are available, as well as for fainter systems, including the increasing population of ambiguous ultra-faint objects that elude spectroscopic observation because they contain almost no red giant branch stars.

\item We take advantage of the high-quality CMD, where the binary sequence is visible with unprecedented clarity, to infer the fraction of binary systems composed of MS stars. To this end, we introduced a new pseudo-CMD, which we dubbed \textit{Binary map}, in which the pseudo-color serves as an indicator of the mass ratio. From this analysis, we derived a binary fraction of $f_{\rm bin}^{q>0.4} = 0.20 \pm 0.02$ for systems with mass ratio $q > 0.4$, and inferred an approximately flat mass-ratio distribution with a slope of $-0.30 \pm 0.15$. By assuming a flat mass-ratio distribution, we extrapolated a total binary fraction of 0.30$\pm$0.03, comparable to what was inferred by \citet{gennaro2018} to characterize the initial mass function of Bo\"otes\,I using ACS/WFC data ($\sim$0.28), and to what was derived in Leo II ($\sim0.30$) and in the inner region of Ursa Minor ($\sim0.33$) \citep{spencer2017,qiu2025}.
The binary fraction is a crucial parameter for constraining the dynamical mass of UFDs, and hence their dark matter content. Indeed, in UFDs, which exhibit relatively small velocity dispersions \citep[e.g.,][]{simon2007}, the contribution from binary stars can reach a few km\,s$^{-1}$, comparable to the observed dispersions in these ancient stellar systems \citep[e.g.,][]{spencer2018,pianta2022}. A relatively large binary fraction of $\sim30\%$, as derived for Bo{\"o}tes\,I, can artificially broaden the measured velocity distribution and thus lead to an incorrect estimate of the dark matter content, as recently demonstrated by studies based on analytic method \citep{gration2025}.   
The binary fraction observed in Bo{\"o}tes\,I is comparable to that typically found in stellar systems with similar stellar mass, such as Galactic and Magellanic Cloud open clusters \citep{cordoni2023,mohandasan2024}, as well as in
the Galactic and Small Magellanic Cloud field for stars of similar masses \citep{offner2023,legnardi2025}. When compared with GCs, Bo{\"o}tes\,I follows the well-known anticorrelation between the binary fraction and the mass of the host stellar system. Using the stellar mass \citep{martin2008}, the dwarf galaxy aligns remarkably well with this relation; however, this agreement breaks down when the dynamical mass is considered \citep{jenkins2021}. Although caution must be exercised when comparing GCs and UFDs, given their different structural properties and evolutionary histories, the fact that Bo{\"o}tes\,I follows the GC trend indicates that dark matter, consistent with its collisionless nature, shapes the global gravitational potential but does not engage in local dynamical interactions that would alter binary fractions, a result that agrees with the simulation run by \cite{livernois2023}, who found that close binaries, as analyzed here, were not affected by the presence of dark matter.

\end{itemize}

\begin{acknowledgements}
We would like to thank the anonymous referee for helpful comments that significantly improved the quality of the paper.
This work has been funded by the European Union – NextGenerationEU RRF M4C2 1.1 (PRIN 2022 2022MMEB9W: "Understanding the formation of globular clusters with their multiple stellar generations", CUP C53D23001200006), and from the European Union’s Horizon 2020 research and innovation programme under the Marie Skłodowska-Curie Grant Agreement No. 101034319 and from the European Union – NextGenerationEU (beneficiary: T. Ziliotto).
This research is based on observations made with the NASA/ESA Hubble Space Telescope obtained from the Space Telescope Science Institute, which is operated by the Association of Universities for Research in Astronomy, Inc., under NASA contract NAS 5–26555. These observations are associated with programs 12549,15317.
This work is based [in part] on observations made with the NASA/ESA/CSA James Webb Space Telescope. The data were obtained from the Mikulski Archive for Space Telescopes at the Space Telescope Science Institute, which is operated by the Association of Universities for Research in Astronomy, Inc., under NASA contract NAS 5-03127 for JWST. These observations are associated with program 3849.
\end{acknowledgements}

%
%

\bibliographystyle{aa}
\bibliography{testbib}

@string{aap = {A\&A}}

@string{aj = {AJ}}

@string{an = {Astron. Nachr.}}

@string{apj = {ApJ}}

@string{apjl = {ApJL}}

@string{apjs = {ApJS}}

@string{araa = {ARA\&A}}

@string{mnras = {MNRAS}}

@string{nature = {Nature}}

@string{pasa = {PASA}}

@string{pasp = {PASP}}

@MISC{anderson2022a,
       author = {{Anderson}, Jay},
        title = "{One-Pass HST Photometry with hst1pass}",
 howpublished = {Instrument Science Report WFC3 2022-5, 55 pages},
         year = 2022,
        month = jul,
        pages = {5},
       adsurl = {https://ui.adsabs.harvard.edu/abs/2022wfc..rept....5A}
}

@ARTICLE{Baumgardt2020,
       author = {{Baumgardt}, H. and {Sollima}, A. and {Hilker}, M.},
        title = "{Absolute V-band magnitudes and mass-to-light ratios of Galactic globular clusters}",
      journal = {\pasa},
         year = 2020,
        month = nov,
       volume = {37},
          eid = {e046},
        pages = {e046},
          doi = {10.1017/pasa.2020.38},
archivePrefix = {arXiv},
       eprint = {2009.09611},
 primaryClass = {astro-ph.GA},
       adsurl = {https://ui.adsabs.harvard.edu/abs/2020PASA...37...46B}
}

@ARTICLE{bellini2017a,
       author = {{Bellini}, A. and {Anderson}, J. and {Bedin}, L.~R. and {King}, I.~R. and {van der Marel}, R.~P. and {Piotto}, G. and {Cool}, A.},
        title = "{The State-of-the-art HST Astro-photometric Analysis of the Core of {\ensuremath{\omega}} Centauri. I. The Catalog}",
      journal = {\apj},
         year = 2017,
        month = jun,
       volume = {842},
       number = {1},
          eid = {6},
        pages = {6},
          doi = {10.3847/1538-4357/aa7059},
archivePrefix = {arXiv},
       eprint = {1704.07425},
 primaryClass = {astro-ph.SR},
       adsurl = {https://ui.adsabs.harvard.edu/abs/2017ApJ...842....6B}
}

@ARTICLE{belokurov2006,
       author = {{Belokurov}, V. and {Zucker}, D.~B. and {Evans}, N.~W. and {Wilkinson}, M.~I. and {Irwin}, M.~J. and {Hodgkin}, S. and {Bramich}, D.~M. and {Irwin}, J.~M. and {Gilmore}, G. and {Willman}, B. and {Vidrih}, S. and {Newberg}, H.~J. and {Wyse}, R.~F.~G. and {Fellhauer}, M. and {Hewett}, P.~C. and {Cole}, N. and {Bell}, E.~F. and {Beers}, T.~C. and {Rockosi}, C.~M. and {Yanny}, B. and {Grebel}, E.~K. and {Schneider}, D.~P. and {Lupton}, R. and {Barentine}, J.~C. and {Brewington}, H. and {Brinkmann}, J. and {Harvanek}, M. and {Kleinman}, S.~J. and {Krzesinski}, J. and {Long}, D. and {Nitta}, A. and {Smith}, J.~A. and {Snedden}, S.~A.},
        title = "{A Faint New Milky Way Satellite in Bootes}",
      journal = {\apjl},
         year = 2006,
        month = aug,
       volume = {647},
       number = {2},
        pages = {L111-L114},
          doi = {10.1086/507324},
archivePrefix = {arXiv},
       eprint = {astro-ph/0604355},
 primaryClass = {astro-ph},
       adsurl = {https://ui.adsabs.harvard.edu/abs/2006ApJ...647L.111B}
}

@ARTICLE{belokurov2010a,
       author = {{Belokurov}, V. and {Walker}, M.~G. and {Evans}, N.~W. and {Gilmore}, G. and {Irwin}, M.~J. and {Just}, D. and {Koposov}, S. and {Mateo}, M. and {Olszewski}, E. and {Watkins}, L. and {Wyrzykowski}, L.},
        title = "{Big Fish, Little Fish: Two New Ultra-faint Satellites of the Milky Way}",
      journal = {\apjl},
         year = 2010,
        month = mar,
       volume = {712},
       number = {1},
        pages = {L103-L106},
          doi = {10.1088/2041-8205/712/1/L103},
archivePrefix = {arXiv},
       eprint = {1002.0504},
 primaryClass = {astro-ph.GA},
       adsurl = {https://ui.adsabs.harvard.edu/abs/2010ApJ...712L.103B}
}

@ARTICLE{cordoni2023,
       author = {{Cordoni}, Giacomo and {Milone}, Antonino P. and {Marino}, Anna F. and {Vesperini}, Enrico and {Dondoglio}, Emanuele and {Legnardi}, Maria Vittoria and {Mohandasan}, Anjana and {Carlos}, Marilia and {Lagioia}, Edoardo P. and {Jang}, Sohee and {Ziliotto}, Tuila},
        title = "{Photometric binaries, mass functions, and structural parameters of 78 Galactic open clusters}",
      journal = {\aap},
         year = 2023,
        month = apr,
       volume = {672},
          eid = {A29},
        pages = {A29},
          doi = {10.1051/0004-6361/202245457},
archivePrefix = {arXiv},
       eprint = {2302.03685},
 primaryClass = {astro-ph.SR},
       adsurl = {https://ui.adsabs.harvard.edu/abs/2023A&A...672A..29C}
}

@ARTICLE{frebel2016,
       author = {{Frebel}, Anna and {Norris}, John E. and {Gilmore}, Gerard and {Wyse}, Rosemary F.~G.},
        title = "{The Chemical Evolution of the Bootes I Ultra-faint Dwarf Galaxy}",
      journal = {\apj},
         year = 2016,
        month = aug,
       volume = {826},
       number = {2},
          eid = {110},
        pages = {110},
          doi = {10.3847/0004-637X/826/2/110},
archivePrefix = {arXiv},
       eprint = {1605.05732},
 primaryClass = {astro-ph.GA},
       adsurl = {https://ui.adsabs.harvard.edu/abs/2016ApJ...826..110F}
}

@ARTICLE{jenkins2021,
       author = {{Jenkins}, Sydney A. and {Li}, Ting S. and {Pace}, Andrew B. and {Ji}, Alexander P. and {Koposov}, Sergey E. and {Mutlu-Pakdil}, Bur{\c{c}}in},
        title = "{Very Large Telescope Spectroscopy of Ultra-faint Dwarf Galaxies. I. Bo{\"o}tes I, Leo IV, and Leo V}",
      journal = {\apj},
         year = 2021,
        month = oct,
       volume = {920},
       number = {2},
          eid = {92},
        pages = {92},
          doi = {10.3847/1538-4357/ac1353},
archivePrefix = {arXiv},
       eprint = {2101.00013},
 primaryClass = {astro-ph.GA},
       adsurl = {https://ui.adsabs.harvard.edu/abs/2021ApJ...920...92J}
}

@ARTICLE{kirby2008,
       author = {{Kirby}, Evan N. and {Guhathakurta}, Puragra and {Sneden}, Christopher},
        title = "{Metallicity and Alpha-Element Abundance Measurement in Red Giant Stars from Medium-Resolution Spectra}",
      journal = {\apj},
         year = 2008,
        month = aug,
       volume = {682},
       number = {2},
        pages = {1217-1233},
          doi = {10.1086/589627},
archivePrefix = {arXiv},
       eprint = {0804.3590},
 primaryClass = {astro-ph},
       adsurl = {https://ui.adsabs.harvard.edu/abs/2008ApJ...682.1217K}
}

@ARTICLE{koposov2011,
       author = {{Koposov}, Sergey E. and {Gilmore}, G. and {Walker}, M.~G. and {Belokurov}, V. and {Evans}, N. Wyn and {Fellhauer}, M. and {Gieren}, W. and {Geisler}, D. and {Monaco}, L. and {Norris}, J.~E. and {Okamoto}, S. and {Pe{\~n}arrubia}, J. and {Wilkinson}, M. and {Wyse}, R.~F.~G. and {Zucker}, D.~B.},
        title = "{Accurate Stellar Kinematics at Faint Magnitudes: Application to the Bo{\"o}tes I Dwarf Spheroidal Galaxy}",
      journal = {\apj},
         year = 2011,
        month = aug,
       volume = {736},
       number = {2},
          eid = {146},
        pages = {146},
          doi = {10.1088/0004-637X/736/2/146},
archivePrefix = {arXiv},
       eprint = {1105.4102},
 primaryClass = {astro-ph.GA},
       adsurl = {https://ui.adsabs.harvard.edu/abs/2011ApJ...736..146K}
}

@ARTICLE{lai2011,
       author = {{Lai}, David K. and {Lee}, Young Sun and {Bolte}, Michael and {Lucatello}, Sara and {Beers}, Timothy C. and {Johnson}, Jennifer A. and {Sivarani}, Thirupathi and {Rockosi}, Constance M.},
        title = "{The [Fe/H], [C/Fe], and [{\ensuremath{\alpha}}/Fe] Distributions of the Bo{\"o}tes I Dwarf Spheroidal Galaxy}",
      journal = {\apj},
         year = 2011,
        month = sep,
       volume = {738},
       number = {1},
          eid = {51},
        pages = {51},
          doi = {10.1088/0004-637X/738/1/51},
archivePrefix = {arXiv},
       eprint = {1106.2168},
 primaryClass = {astro-ph.GA},
       adsurl = {https://ui.adsabs.harvard.edu/abs/2011ApJ...738...51L}
}

@ARTICLE{martin2008,
       author = {{Martin}, Nicolas F. and {de Jong}, Jelte T.~A. and {Rix}, Hans-Walter},
        title = "{A Comprehensive Maximum Likelihood Analysis of the Structural Properties of Faint Milky Way Satellites}",
      journal = {\apj},
         year = 2008,
        month = sep,
       volume = {684},
       number = {2},
        pages = {1075-1092},
          doi = {10.1086/590336},
archivePrefix = {arXiv},
       eprint = {0805.2945},
 primaryClass = {astro-ph},
       adsurl = {https://ui.adsabs.harvard.edu/abs/2008ApJ...684.1075M}
}

@ARTICLE{kim2015a,
       author = {{Kim}, Dongwon and {Jerjen}, Helmut and {Mackey}, Dougal and {Da Costa}, Gary S. and {Milone}, Antonino P.},
        title = "{A Hero{\textquoteright}s Dark Horse: Discovery of an Ultra-faint Milky Way Satellite in Pegasus}",
      journal = {\apjl},
         year = 2015,
        month = may,
       volume = {804},
       number = {2},
          eid = {L44},
        pages = {L44},
          doi = {10.1088/2041-8205/804/2/L44},
archivePrefix = {arXiv},
       eprint = {1503.08268},
 primaryClass = {astro-ph.GA},
       adsurl = {https://ui.adsabs.harvard.edu/abs/2015ApJ...804L..44K}
}

@ARTICLE{legnardi2022,
       author = {{Legnardi}, M.~V. and {Milone}, A.~P. and {Armillotta}, L. and {Marino}, A.~F. and {Cordoni}, G. and {Renzini}, A. and {Vesperini}, E. and {D'Antona}, F. and {McKenzie}, M. and {Yong}, D. and {Dondoglio}, E. and {Lagioia}, E.~P. and {Carlos}, M. and {Tailo}, M. and {Jang}, S. and {Mohandasan}, A.},
        title = "{Constraining the original composition of the gas forming first-generation stars in globular clusters}",
      journal = {\mnras},
         year = 2022,
        month = jun,
       volume = {513},
       number = {1},
        pages = {735-751},
          doi = {10.1093/mnras/stac734},
archivePrefix = {arXiv},
       eprint = {2203.07571},
 primaryClass = {astro-ph.GA},
       adsurl = {https://ui.adsabs.harvard.edu/abs/2022MNRAS.513..735L}
}

@ARTICLE{legnardi2024,
       author = {{Legnardi}, M.~V. and {Milone}, A.~P. and {Cordoni}, G. and {Marino}, A.~F. and {Dondoglio}, E. and {Jang}, S. and {Lagioia}, E.~P. and {Muratore}, F. and {Ziliotto}, T. and {Bortolan}, E. and {Mohandasan}, A.},
        title = "{The original composition of the gas forming first-generation stars in clusters: Insights from HST and JWST}",
      journal = {\aap},
         year = 2024,
        month = jul,
       volume = {687},
          eid = {A160},
        pages = {A160},
          doi = {10.1051/0004-6361/202449533},
archivePrefix = {arXiv},
       eprint = {2405.02006},
 primaryClass = {astro-ph.GA},
       adsurl = {https://ui.adsabs.harvard.edu/abs/2024A&A...687A.160L}
}

@ARTICLE{milone2012,
       author = {{Milone}, A.~P. and {Piotto}, G. and {Bedin}, L.~R. and {Aparicio}, A. and {Anderson}, J. and {Sarajedini}, A. and {Marino}, A.~F. and {Moretti}, A. and {Davies}, M.~B. and {Chaboyer}, B. and {Dotter}, A. and {Hempel}, M. and {Mar{\'\i}n-Franch}, A. and {Majewski}, S. and {Paust}, N.~E.~Q. and {Reid}, I.~N. and {Rosenberg}, A. and {Siegel}, M.},
        title = "{The ACS survey of Galactic globular clusters. XII. Photometric binaries along the main sequence}",
      journal = {\aap},
         year = 2012,
        month = apr,
       volume = {540},
          eid = {A16},
        pages = {A16},
          doi = {10.1051/0004-6361/201016384},
archivePrefix = {arXiv},
       eprint = {1111.0552},
 primaryClass = {astro-ph.SR},
       adsurl = {https://ui.adsabs.harvard.edu/abs/2012A&A...540A..16M}
}

@ARTICLE{milone2016,
       author = {{Milone}, A.~P. and {Marino}, A.~F. and {Bedin}, L.~R. and {Dotter}, A. and {Jerjen}, H. and {Kim}, D. and {Nardiello}, D. and {Piotto}, G. and {Cong}, J.},
        title = "{The binary populations of eight globular clusters in the outer halo of the Milky Way}",
      journal = {\mnras},
         year = 2016,
        month = jan,
       volume = {455},
       number = {3},
        pages = {3009-3019},
          doi = {10.1093/mnras/stv2415},
archivePrefix = {arXiv},
       eprint = {1510.05086},
 primaryClass = {astro-ph.SR},
       adsurl = {https://ui.adsabs.harvard.edu/abs/2016MNRAS.455.3009M}
}

@ARTICLE{milone2023,
       author = {{Milone}, A.~P. and {Cordoni}, G. and {Marino}, A.~F. and {D'Antona}, F. and {Bellini}, A. and {Di Criscienzo}, M. and {Dondoglio}, E. and {Lagioia}, E.~P. and {Langer}, N. and {Legnardi}, M.~V. and {Libralato}, M. and {Baumgardt}, H. and {Bettinelli}, M. and {Cavecchi}, Y. and {de Grijs}, R. and {Deng}, L. and {Hastings}, B. and {Li}, C. and {Mohandasan}, A. and {Renzini}, A. and {Vesperini}, E. and {Wang}, C. and {Ziliotto}, T. and {Carlos}, M. and {Costa}, G. and {Dell'Agli}, F. and {Di Stefano}, S. and {Jang}, S. and {Martorano}, M. and {Simioni}, M. and {Tailo}, M. and {Ventura}, P.},
        title = "{Hubble Space Telescope survey of Magellanic Cloud star clusters. Photometry and astrometry of 113 clusters and early results}",
      journal = {\aap},
         year = 2023,
        month = apr,
       volume = {672},
          eid = {A161},
        pages = {A161},
          doi = {10.1051/0004-6361/202244798},
archivePrefix = {arXiv},
       eprint = {2212.07978},
 primaryClass = {astro-ph.SR},
       adsurl = {https://ui.adsabs.harvard.edu/abs/2023A&A...672A.161M}
}

@ARTICLE{mohandasan2024,
       author = {{Mohandasan}, Anjana and {Milone}, Antonino P. and {Cordoni}, Giacomo and {Dondoglio}, Emanuele and {Lagioia}, Edoardo P. and {Legnardi}, Maria Vittoria and {Ziliotto}, Tuila and {Jang}, Sohee and {Marino}, Anna F. and {Carlos}, Mar{\'\i}lia},
        title = "{Photometric binaries in 14 Magellanic Cloud star clusters}",
      journal = {\aap},
         year = 2024,
        month = jan,
       volume = {681},
          eid = {A42},
        pages = {A42},
          doi = {10.1051/0004-6361/202347424},
archivePrefix = {arXiv},
       eprint = {2310.15345},
 primaryClass = {astro-ph.SR},
       adsurl = {https://ui.adsabs.harvard.edu/abs/2024A&A...681A..42M}
}

@ARTICLE{norris2008,
       author = {{Norris}, John E. and {Gilmore}, Gerard and {Wyse}, Rosemary F.~G. and {Wilkinson}, Mark I. and {Belokurov}, V. and {Evans}, N. Wyn and {Zucker}, Daniel B.},
        title = "{The Abundance Spread in the Bo{\"o}tes I Dwarf Spheroidal Galaxy}",
      journal = {\apjl},
         year = 2008,
        month = dec,
       volume = {689},
       number = {2},
        pages = {L113},
          doi = {10.1086/595962},
       adsurl = {https://ui.adsabs.harvard.edu/abs/2008ApJ...689L.113N}
}

@ARTICLE{norris2010a,
       author = {{Norris}, John E. and {Wyse}, Rosemary F.~G. and {Gilmore}, Gerard and {Yong}, David and {Frebel}, Anna and {Wilkinson}, Mark I. and {Belokurov}, V. and {Zucker}, Daniel B.},
        title = "{Chemical Enrichment in the Faintest Galaxies: The Carbon and Iron Abundance Spreads in the Bo{\"o}tes I Dwarf Spheroidal Galaxy and the Segue 1 System}",
      journal = {\apj},
         year = 2010,
        month = nov,
       volume = {723},
       number = {2},
        pages = {1632-1650},
          doi = {10.1088/0004-637X/723/2/1632},
archivePrefix = {arXiv},
       eprint = {1008.0137},
 primaryClass = {astro-ph.GA},
       adsurl = {https://ui.adsabs.harvard.edu/abs/2010ApJ...723.1632N}
}

@ARTICLE{norris2010b,
       author = {{Norris}, John E. and {Yong}, David and {Gilmore}, Gerard and {Wyse}, Rosemary F.~G.},
        title = "{Boo-1137{\textemdash}an Extremely Metal-Poor Star in the Ultra-Faint Dwarf Spheroidal Galaxy Bo{\"o}tes I}",
      journal = {\apj},
         year = 2010,
        month = mar,
       volume = {711},
       number = {1},
        pages = {350-360},
          doi = {10.1088/0004-637X/711/1/350},
archivePrefix = {arXiv},
       eprint = {0911.5350},
 primaryClass = {astro-ph.GA},
       adsurl = {https://ui.adsabs.harvard.edu/abs/2010ApJ...711..350N}
}

@ARTICLE{okamoto2012,
       author = {{Okamoto}, Sakurako and {Arimoto}, Nobuo and {Yamada}, Yoshihiko and {Onodera}, Masato},
        title = "{Stellar Populations and Structural Properties of Ultra Faint Dwarf Galaxies, Canes Venatici I, Bo{\"o}tes I, Canes Venatici II, and Leo IV}",
      journal = {\apj},
         year = 2012,
        month = jan,
       volume = {744},
       number = {2},
          eid = {96},
        pages = {96},
          doi = {10.1088/0004-637X/744/2/96},
archivePrefix = {arXiv},
       eprint = {1110.2882},
 primaryClass = {astro-ph.GA},
       adsurl = {https://ui.adsabs.harvard.edu/abs/2012ApJ...744...96O}
}

@ARTICLE{simon2019,
       author = {{Simon}, Joshua D.},
        title = "{The Faintest Dwarf Galaxies}",
      journal = {\araa},
         year = 2019,
        month = aug,
       volume = {57},
        pages = {375-415},
          doi = {10.1146/annurev-astro-091918-104453},
archivePrefix = {arXiv},
       eprint = {1901.05465},
 primaryClass = {astro-ph.GA},
       adsurl = {https://ui.adsabs.harvard.edu/abs/2019ARA&A..57..375S}
}

@ARTICLE{spencer2018,
       author = {{Spencer}, Meghin E. and {Mateo}, Mario and {Olszewski}, Edward W. and {Walker}, Matthew G. and {McConnachie}, Alan W. and {Kirby}, Evan N.},
        title = "{The Binary Fraction of Stars in Dwarf Galaxies: The Cases of Draco and Ursa Minor}",
      journal = {\aj},
         year = 2018,
        month = dec,
       volume = {156},
       number = {6},
          eid = {257},
        pages = {257},
          doi = {10.3847/1538-3881/aae3e4},
archivePrefix = {arXiv},
       eprint = {1811.06597},
 primaryClass = {astro-ph.GA},
       adsurl = {https://ui.adsabs.harvard.edu/abs/2018AJ....156..257S}
}

@ARTICLE{willman2011,
       author = {{Willman}, Beth and {Geha}, Marla and {Strader}, Jay and {Strigari}, Louis E. and {Simon}, Joshua D. and {Kirby}, Evan and {Ho}, Nhung and {Warres}, Alex},
        title = "{Willman 1{\textemdash}A Probable Dwarf Galaxy with an Irregular Kinematic Distribution}",
      journal = {\aj},
         year = 2011,
        month = oct,
       volume = {142},
       number = {4},
          eid = {128},
        pages = {128},
          doi = {10.1088/0004-6256/142/4/128},
archivePrefix = {arXiv},
       eprint = {1007.3499},
 primaryClass = {astro-ph.GA},
       adsurl = {https://ui.adsabs.harvard.edu/abs/2011AJ....142..128W}
}

@ARTICLE{simon2007,
       author = {{Simon}, Joshua D. and {Geha}, Marla},
        title = "{The Kinematics of the Ultra-faint Milky Way Satellites: Solving the Missing Satellite Problem}",
      journal = {\apj},
         year = 2007,
        month = nov,
       volume = {670},
       number = {1},
        pages = {313-331},
          doi = {10.1086/521816},
archivePrefix = {arXiv},
       eprint = {0706.0516},
 primaryClass = {astro-ph},
       adsurl = {https://ui.adsabs.harvard.edu/abs/2007ApJ...670..313S}
}

@ARTICLE{gration2025,
       author = {{Gration}, A. and {Hendriks}, D.~D. and {Das}, P. and {Heber}, D. and {Izzard}, R.~G.},
        title = "{Stellar velocity distributions in binary-rich ultrafaint dwarf galaxies}",
      journal = {\mnras},
         year = 2025,
        month = oct,
       volume = {543},
       number = {2},
        pages = {1120-1132},
          doi = {10.1093/mnras/staf1481},
archivePrefix = {arXiv},
       eprint = {2509.14316},
 primaryClass = {astro-ph.GA},
       adsurl = {https://ui.adsabs.harvard.edu/abs/2025MNRAS.543.1120G}
}

@ARTICLE{pianta2022,
       author = {{Pianta}, Camilla and {Capuzzo-Dolcetta}, Roberto and {Carraro}, Giovanni},
        title = "{The Impact of Binaries on the Dynamical Mass Estimate of Dwarf Galaxies}",
      journal = {\apj},
         year = 2022,
        month = nov,
       volume = {939},
       number = {1},
          eid = {3},
        pages = {3},
          doi = {10.3847/1538-4357/ac9303},
archivePrefix = {arXiv},
       eprint = {2209.08296},
 primaryClass = {astro-ph.GA},
       adsurl = {https://ui.adsabs.harvard.edu/abs/2022ApJ...939....3P}
}

@ARTICLE{muratore2024,
       author = {{Muratore}, F. and {Milone}, A.~P. and {D'Antona}, F. and {Nastasio}, E.~J. and {Cordoni}, G. and {Legnardi}, M.~V. and {He}, C. and {Ziliotto}, T. and {Dondoglio}, E. and {Bernizzoni}, M. and et al.},
        title = "{Hubble Space Telescope survey of Magellanic Cloud star clusters. Binaries among the split main sequences of NGC 1818, NGC 1850, and NGC 2164}",
      journal = {\aap},
         year = 2024,
        month = dec,
       volume = {692},
          eid = {A135},
        pages = {A135},
          doi = {10.1051/0004-6361/202451310},
archivePrefix = {arXiv},
       eprint = {2411.02508},
 primaryClass = {astro-ph.SR},
       adsurl = {https://ui.adsabs.harvard.edu/abs/2024A&A...692A.135M}
}

@ARTICLE{legnardi2025,
       author = {{Legnardi}, M.~V. and {Muratore}, F. and {Milone}, A.~P. and {Cordoni}, G. and {Ziliotto}, T. and {Dondoglio}, E. and {Marino}, A.~F. and {Mastrobuono-Battisti}, A. and {Bortolan}, E. and {Lagioia}, E.~P. and {Tailo}, M.},
        title = "{The Small Magellanic Cloud through the lens of the James Webb Space Telescope: Binaries and the mass function in the galaxy's outskirts}",
      journal = {\aap},
         year = 2025,
        month = oct,
       volume = {702},
          eid = {A180},
        pages = {A180},
          doi = {10.1051/0004-6361/202556239},
archivePrefix = {arXiv},
       eprint = {2509.08687},
 primaryClass = {astro-ph.GA},
       adsurl = {https://ui.adsabs.harvard.edu/abs/2025A&A...702A.180L}
}

@INPROCEEDINGS{offner2023,
       author = {{Offner}, S.~S.~R. and {Moe}, M. and {Kratter}, K.~M. and {Sadavoy}, S.~I. and {Jensen}, E.~L.~N. and {Tobin}, J.~J.},
        title = "{The Origin and Evolution of Multiple Star Systems}",
    booktitle = {Protostars and Planets VII},
         year = 2023,
       editor = {{Inutsuka}, S. and {Aikawa}, Y. and {Muto}, T. and {Tomida}, K. and {Tamura}, M.},
       series = {Astronomical Society of the Pacific Conference Series},
       volume = {534},
        month = jul,
        pages = {275},
          doi = {10.48550/arXiv.2203.10066},
archivePrefix = {arXiv},
       eprint = {2203.10066},
 primaryClass = {astro-ph.SR},
       adsurl = {https://ui.adsabs.harvard.edu/abs/2023ASPC..534..275O}
}

@ARTICLE{spencer2017,
       author = {{Spencer}, Meghin E. and {Mateo}, Mario and {Walker}, Matthew G. and {Olszewski}, Edward W. and {McConnachie}, Alan W. and {Kirby}, Evan N. and {Koch}, Andreas},
        title = "{The Binary Fraction of Stars in Dwarf Galaxies: The Case of Leo II}",
      journal = {\aj},
         year = 2017,
        month = jun,
       volume = {153},
       number = {6},
          eid = {254},
        pages = {254},
          doi = {10.3847/1538-3881/aa6d51},
archivePrefix = {arXiv},
       eprint = {1706.04184},
 primaryClass = {astro-ph.GA},
       adsurl = {https://ui.adsabs.harvard.edu/abs/2017AJ....153..254S}
}

@ARTICLE{mcconnachie2010,
       author = {{McConnachie}, Alan W. and {C{\^o}t{\'e}}, Patrick},
        title = "{Revisiting the Influence of Unidentified Binaries on Velocity Dispersion Measurements in Ultra-faint Stellar Systems}",
      journal = {\apjl},
         year = 2010,
        month = oct,
       volume = {722},
       number = {2},
        pages = {L209-L214},
          doi = {10.1088/2041-8205/722/2/L209},
archivePrefix = {arXiv},
       eprint = {1009.4205},
 primaryClass = {astro-ph.CO},
       adsurl = {https://ui.adsabs.harvard.edu/abs/2010ApJ...722L.209M}
}

@ARTICLE{anderson2008,
       author = {{Anderson}, Jay and {Sarajedini}, Ata and {Bedin}, Luigi R. and {King}, Ivan R. and {Piotto}, Giampaolo and {Reid}, I. Neill and {Siegel}, Michael and {Majewski}, Steven R. and {Paust}, Nathaniel E.~Q. and {Aparicio}, Antonio and {Milone}, Antonino P. and {Chaboyer}, Brian and {Rosenberg}, Alfred},
        title = "{The Acs Survey of Globular Clusters. V. Generating a Comprehensive Star Catalog for each Cluster}",
      journal = {\aj},
         year = 2008,
        month = jun,
       volume = {135},
       number = {6},
        pages = {2055-2073},
          doi = {10.1088/0004-6256/135/6/2055},
archivePrefix = {arXiv},
       eprint = {0804.2025},
 primaryClass = {astro-ph},
       adsurl = {https://ui.adsabs.harvard.edu/abs/2008AJ....135.2055A}
}

@ARTICLE{mcconnachie2012,
       author = {{McConnachie}, Alan W.},
        title = "{The Observed Properties of Dwarf Galaxies in and around the Local Group}",
      journal = {\aj},
         year = 2012,
        month = jul,
       volume = {144},
       number = {1},
          eid = {4},
        pages = {4},
          doi = {10.1088/0004-6256/144/1/4},
archivePrefix = {arXiv},
       eprint = {1204.1562},
 primaryClass = {astro-ph.CO},
       adsurl = {https://ui.adsabs.harvard.edu/abs/2012AJ....144....4M}
}

@ARTICLE{brown2014,
       author = {{Brown}, Thomas M. and {Tumlinson}, Jason and {Geha}, Marla and {Simon}, Joshua D. and {Vargas}, Luis C. and {VandenBerg}, Don A. and {Kirby}, Evan N. and {Kalirai}, Jason S. and {Avila}, Roberto J. and {Gennaro}, Mario and et al.},
        title = "{The Quenching of the Ultra-faint Dwarf Galaxies in the Reionization Era}",
      journal = {\apj},
         year = 2014,
        month = dec,
       volume = {796},
       number = {2},
          eid = {91},
        pages = {91},
          doi = {10.1088/0004-637X/796/2/91},
archivePrefix = {arXiv},
       eprint = {1410.0681},
 primaryClass = {astro-ph.GA},
       adsurl = {https://ui.adsabs.harvard.edu/abs/2014ApJ...796...91B}
}

@ARTICLE{battaglia2022,
       author = {{Battaglia}, Giuseppina and {Nipoti}, Carlo},
        title = "{Publisher Correction: Stellar dynamics and dark matter in Local Group dwarf galaxies}",
      journal = {Nature Astronomy},
         year = 2022,
        month = dec,
       volume = {6},
        pages = {1492-1492},
          doi = {10.1038/s41550-022-01829-2},
       adsurl = {https://ui.adsabs.harvard.edu/abs/2022NatAs...6.1492B}
}

@ARTICLE{milone2017,
       author = {{Milone}, A.~P. and {Piotto}, G. and {Renzini}, A. and {Marino}, A.~F. and {Bedin}, L.~R. and {Vesperini}, E. and {D'Antona}, F. and {Nardiello}, D. and {Anderson}, J. and {King}, I.~R. and et al.},
        title = "{The Hubble Space Telescope UV Legacy Survey of Galactic globular clusters - IX. The Atlas of multiple stellar populations}",
      journal = {\mnras},
         year = 2017,
        month = jan,
       volume = {464},
       number = {3},
        pages = {3636-3656},
          doi = {10.1093/mnras/stw2531},
archivePrefix = {arXiv},
       eprint = {1610.00451},
 primaryClass = {astro-ph.SR},
       adsurl = {https://ui.adsabs.harvard.edu/abs/2017MNRAS.464.3636M}
}

@ARTICLE{milone2025a,
       author = {{Milone}, A.~P. and {Marino}, A.~F. and {Bernizzoni}, M. and {Muratore}, F. and {Legnardi}, M.~V. and {Barbieri}, M. and {Bortolan}, E. and {Bouras}, A. and {Bruce}, J. and {Cordoni}, G. and et al.},
        title = "{A JWST project on 47 Tucanae: Binaries among multiple populations}",
      journal = {\aap},
         year = 2025,
        month = jun,
       volume = {698},
          eid = {A247},
        pages = {A247},
          doi = {10.1051/0004-6361/202452136},
archivePrefix = {arXiv},
       eprint = {2503.19214},
 primaryClass = {astro-ph.SR},
       adsurl = {https://ui.adsabs.harvard.edu/abs/2025A&A...698A.247M}
}

@ARTICLE{hidalgo2018a,
       author = {{Hidalgo}, Sebastian L. and {Pietrinferni}, Adriano and {Cassisi}, Santi and {Salaris}, Maurizio and {Mucciarelli}, Alessio and {Savino}, Alessandro and {Aparicio}, Antonio and {Silva Aguirre}, Victor and {Verma}, Kuldeep},
        title = "{The Updated BaSTI Stellar Evolution Models and Isochrones. I. Solar-scaled Calculations}",
      journal = {\apj},
         year = 2018,
        month = apr,
       volume = {856},
       number = {2},
          eid = {125},
        pages = {125},
          doi = {10.3847/1538-4357/aab158},
archivePrefix = {arXiv},
       eprint = {1802.07319},
 primaryClass = {astro-ph.GA},
       adsurl = {https://ui.adsabs.harvard.edu/abs/2018ApJ...856..125H}
}

@ARTICLE{cignoni2010a,
       author = {{Cignoni}, Michele and {Tosi}, Monica},
        title = "{Star Formation Histories of Dwarf Galaxies from the Colour-Magnitude Diagrams of Their Resolved Stellar Populations}",
      journal = {Advances in Astronomy},
         year = 2010,
        month = jan,
       volume = {2010},
          eid = {158568},
        pages = {158568},
          doi = {10.1155/2010/158568},
archivePrefix = {arXiv},
       eprint = {0909.4234},
 primaryClass = {astro-ph.GA},
       adsurl = {https://ui.adsabs.harvard.edu/abs/2010AdAst2010E...3C}
}

@ARTICLE{cordoni2022a,
       author = {{Cordoni}, Giacomo and {Milone}, Antonino P. and {Marino}, Anna F. and {Cignoni}, Michele and {Lagioia}, Edoardo P. and {Tailo}, Marco and {Carlos}, Mar{\'\i}lia and {Dondoglio}, Emanuele and {Jang}, Sohee and {Mohandasan}, Anjana and et al.},
        title = "{NGC1818 unveils the origin of the extended main-sequence turn-off in young Magellanic Clouds clusters}",
      journal = {Nature Communications},
         year = 2022,
        month = jul,
       volume = {13},
          eid = {4325},
        pages = {4325},
          doi = {10.1038/s41467-022-31977-y},
archivePrefix = {arXiv},
       eprint = {2208.09467},
 primaryClass = {astro-ph.SR},
       adsurl = {https://ui.adsabs.harvard.edu/abs/2022NatCo..13.4325C}
}

@ARTICLE{harris1996a,
       author = {{Harris}, William E.},
        title = "{A Catalog of Parameters for Globular Clusters in the Milky Way}",
      journal = {\aj},
         year = 1996,
        month = oct,
       volume = {112},
        pages = {1487},
          doi = {10.1086/118116},
       adsurl = {https://ui.adsabs.harvard.edu/abs/1996AJ....112.1487H}
}

@ARTICLE{stauffer1980,
       author = {{Stauffer}, J.~R.},
        title = "{Observations of Pre-main-sequence stars in the Pleiades.}",
      journal = {\aj},
         year = 1980,
        month = oct,
       volume = {85},
        pages = {1341-1353},
          doi = {10.1086/112805},
       adsurl = {https://ui.adsabs.harvard.edu/abs/1980AJ.....85.1341S}
}

@ARTICLE{gennaro2018,
       author = {{Gennaro}, Mario and {Tchernyshyov}, Kirill and {Brown}, Thomas M. and {Geha}, Marla and {Avila}, Roberto J. and {Guhathakurta}, Puragra and {Kalirai}, Jason S. and {Kirby}, Evan N. and {Renzini}, Alvio and {Simon}, Joshua D. and {Tumlinson}, Jason and {Vargas}, Luis C.},
        title = "{Evidence of a Non-universal Stellar Initial Mass Function. Insights from HST Optical Imaging of Six Ultra-faint Dwarf Milky Way Satellites}",
      journal = {\apj},
         year = 2018,
        month = mar,
       volume = {855},
       number = {1},
          eid = {20},
        pages = {20},
          doi = {10.3847/1538-4357/aaa973},
archivePrefix = {arXiv},
       eprint = {1801.06195},
 primaryClass = {astro-ph.GA},
       adsurl = {https://ui.adsabs.harvard.edu/abs/2018ApJ...855...20G}
}

@ARTICLE{anderson2000,
       author = {{Anderson}, Jay and {King}, Ivan R.},
        title = "{Toward High-Precision Astrometry with WFPC2. I. Deriving an Accurate Point-Spread Function}",
      journal = {\pasp},
         year = 2000,
        month = oct,
       volume = {112},
       number = {776},
        pages = {1360-1382},
          doi = {10.1086/316632},
archivePrefix = {arXiv},
       eprint = {astro-ph/0006325},
 primaryClass = {astro-ph},
       adsurl = {https://ui.adsabs.harvard.edu/abs/2000PASP..112.1360A}
}

@ARTICLE{sabbi2016a,
       author = {{Sabbi}, E. and {Lennon}, D.~J. and {Anderson}, J. and {Cignoni}, M. and {van der Marel}, R.~P. and {Zaritsky}, D. and {De Marchi}, G. and {Panagia}, N. and {Gouliermis}, D.~A. and {Grebel}, E.~K. and {Gallagher}, III, J.~S. and {Smith}, L.~J. and {Sana}, H. and {Aloisi}, A. and {Tosi}, M. and {Evans}, C.~J. and {Arab}, H. and {Boyer}, M. and {de Mink}, S.~E. and {Gordon}, K. and {Koekemoer}, A.~M. and {Larsen}, S.~S. and {Ryon}, J.~E. and {Zeidler}, P.},
        title = "{Hubble Tarantula Treasury Project. III. Photometric Catalog and Resulting Constraints on the Progression of Star Formation in the 30 Doradus Region}",
      journal = {\apjs},
         year = 2016,
        month = jan,
       volume = {222},
       number = {1},
          eid = {11},
        pages = {11},
          doi = {10.3847/0067-0049/222/1/11},
archivePrefix = {arXiv},
       eprint = {1511.06021},
 primaryClass = {astro-ph.GA},
       adsurl = {https://ui.adsabs.harvard.edu/abs/2016ApJS..222...11S}
}

@ARTICLE{marino2015,
       author = {{Marino}, A.~F. and {Milone}, A.~P. and {Karakas}, A.~I. and {Casagrande}, L. and {Yong}, D. and {Shingles}, L. and {Da Costa}, G. and {Norris}, J.~E. and {Stetson}, P.~B. and {Lind}, K. and et al.},
        title = "{Iron and s-elements abundance variations in NGC 5286: comparison with `anomalous' globular clusters and Milky Way satellites}",
      journal = {\mnras},
         year = 2015,
        month = jun,
       volume = {450},
       number = {1},
        pages = {815-845},
          doi = {10.1093/mnras/stv420},
archivePrefix = {arXiv},
       eprint = {1502.07438},
 primaryClass = {astro-ph.SR},
       adsurl = {https://ui.adsabs.harvard.edu/abs/2015MNRAS.450..815M}
}

@ARTICLE{sacchi2021,
       author = {{Sacchi}, Elena and {Richstein}, Hannah and {Kallivayalil}, Nitya and {van der Marel}, Roeland and {Libralato}, Mattia and {Zivick}, Paul and {Besla}, Gurtina and {Brown}, Thomas M. and {Choi}, Yumi and {Deason}, Alis and et al.},
        title = "{Star Formation Histories of Ultra-faint Dwarf Galaxies: Environmental Differences between Magellanic and Non-Magellanic Satellites?}",
      journal = {\apjl},
         year = 2021,
        month = oct,
       volume = {920},
       number = {1},
          eid = {L19},
        pages = {L19},
          doi = {10.3847/2041-8213/ac2aa3},
archivePrefix = {arXiv},
       eprint = {2108.04271},
 primaryClass = {astro-ph.GA},
       adsurl = {https://ui.adsabs.harvard.edu/abs/2021ApJ...920L..19S}
}

@ARTICLE{savino2023,
       author = {{Savino}, Alessandro and {Weisz}, Daniel R. and {Skillman}, Evan D. and {Dolphin}, Andrew and {Cole}, Andrew A. and {Kallivayalil}, Nitya and {Wetzel}, Andrew and {Anderson}, Jay and {Besla}, Gurtina and {Boylan-Kolchin}, Michael and et al.},
        title = "{The Hubble Space Telescope Survey of M31 Satellite Galaxies. II. The Star Formation Histories of Ultrafaint Dwarf Galaxies}",
      journal = {\apj},
         year = 2023,
        month = oct,
       volume = {956},
       number = {2},
          eid = {86},
        pages = {86},
          doi = {10.3847/1538-4357/acf46f},
archivePrefix = {arXiv},
       eprint = {2305.13360},
 primaryClass = {astro-ph.GA},
       adsurl = {https://ui.adsabs.harvard.edu/abs/2023ApJ...956...86S}
}

@ARTICLE{Durbin2025,
       author = {{Durbin}, Meredith J. and {Choi}, Yumi and {Savino}, Alessandro and {Weisz}, Daniel R. and {Dolphin}, Andrew E. and {Dalcanton}, Julianne J. and {Jeon}, Myoungwon and {Kallivayalil}, Nitya and {Li}, Ting S. and {Pace}, Andrew B. and et al.},
        title = "{The HST Legacy Archival Uniform Reduction of Local Group Imaging (LAURELIN). I. Photometry and Star Formation Histories for 36 Ultra-faint Dwarf Galaxies}",
      journal = {\apj},
         year = 2025,
        month = oct,
       volume = {992},
       number = {1},
          eid = {106},
        pages = {106},
          doi = {10.3847/1538-4357/ae00c8},
archivePrefix = {arXiv},
       eprint = {2505.18252},
 primaryClass = {astro-ph.GA},
       adsurl = {https://ui.adsabs.harvard.edu/abs/2025ApJ...992..106D}
}

@ARTICLE{Zoutendijk2021a,
       author = {{Zoutendijk}, Sebastiaan L. and {Brinchmann}, Jarle and {Bouch{\'e}}, Nicolas F. and {den Brok}, Mark and {Krajnovi{\'c}}, Davor and {Kuijken}, Konrad and {Maseda}, Michael V. and {Schaye}, Joop},
        title = "{The MUSE-Faint survey. II. The dark-matter density profile of the ultra-faint dwarf galaxy Eridanus 2}",
      journal = {\aap},
         year = 2021,
        month = jul,
       volume = {651},
          eid = {A80},
        pages = {A80},
          doi = {10.1051/0004-6361/202040239},
archivePrefix = {arXiv},
       eprint = {2101.00253},
 primaryClass = {astro-ph.GA},
       adsurl = {https://ui.adsabs.harvard.edu/abs/2021A&A...651A..80Z}
}

@ARTICLE{longeard2022,
       author = {{Longeard}, Nicolas and {Jablonka}, Pascale and {Arentsen}, Anke and {Thomas}, Guillaume F. and {Aguado}, David S. and {Carlberg}, Raymond G. and {Lucchesi}, Romain and {Malhan}, Khyati and {Martin}, Nicolas and {McConnachie}, Alan W. and et al.},
        title = "{The Pristine dwarf galaxy survey - IV. Probing the outskirts of the dwarf galaxy Bo{\"o}tes I}",
      journal = {\mnras},
         year = 2022,
        month = oct,
       volume = {516},
       number = {2},
        pages = {2348-2362},
          doi = {10.1093/mnras/stac1827},
archivePrefix = {arXiv},
       eprint = {2107.10849},
 primaryClass = {astro-ph.GA},
       adsurl = {https://ui.adsabs.harvard.edu/abs/2022MNRAS.516.2348L}
}

@ARTICLE{sandford2025,
       author = {{Sandford}, Nathan R. and {Li}, Ting S. and {Koposov}, Sergey E. and {Hayashi}, Kohei and {Pace}, Andrew B. and {Erkal}, Denis and {Bovy}, Jo and {Da Costa}, Gary S. and {Cullinane}, Lara R. and {Ji}, Alexander P. and et al.},
        title = "{Chemodynamics of Bo{\"o}tesI with $S^{5}$: Revised Velocity Gradient, Dark Matter Density, and Galactic Chemical Evolution Constraints}",
      journal = {arXiv e-prints},
         year = 2025,
        month = sep,
          eid = {arXiv:2509.02546},
        pages = {arXiv:2509.02546},
          doi = {10.48550/arXiv.2509.02546},
archivePrefix = {arXiv},
       eprint = {2509.02546},
 primaryClass = {astro-ph.GA},
       adsurl = {https://ui.adsabs.harvard.edu/abs/2025arXiv250902546S}
}

@ARTICLE{gilmore2013,
       author = {{Gilmore}, Gerard and {Norris}, John E. and {Monaco}, Lorenzo and {Yong}, David and {Wyse}, Rosemary F.~G. and {Geisler}, D.},
        title = "{Elemental Abundances and their Implications for the Chemical Enrichment of the Bo{\"o}tes I Ultrafaint Galaxy}",
      journal = {\apj},
         year = 2013,
        month = jan,
       volume = {763},
       number = {1},
          eid = {61},
        pages = {61},
          doi = {10.1088/0004-637X/763/1/61},
archivePrefix = {arXiv},
       eprint = {1212.0598},
 primaryClass = {astro-ph.GA},
       adsurl = {https://ui.adsabs.harvard.edu/abs/2013ApJ...763...61G}
}

@ARTICLE{ishigaki2014,
       author = {{Ishigaki}, M.~N. and {Aoki}, W. and {Arimoto}, N. and {Okamoto}, S.},
        title = "{Chemical compositions of six metal-poor stars in the ultra-faint dwarf spheroidal galaxy Bo{\"o}tes I}",
      journal = {\aap},
         year = 2014,
        month = feb,
       volume = {562},
          eid = {A146},
        pages = {A146},
          doi = {10.1051/0004-6361/201322796},
archivePrefix = {arXiv},
       eprint = {1401.1265},
 primaryClass = {astro-ph.GA},
       adsurl = {https://ui.adsabs.harvard.edu/abs/2014A&A...562A.146I}
}

@ARTICLE{milone2015,
       author = {{Milone}, A.~P. and {Marino}, A.~F. and {Piotto}, G. and {Renzini}, A. and {Bedin}, L.~R. and {Anderson}, J. and {Cassisi}, S. and {D'Antona}, F. and {Bellini}, A. and {Jerjen}, H. and et al.},
        title = "{The Hubble Space Telescope UV Legacy Survey of Galactic Globular Clusters. III. A Quintuple Stellar Population in NGC 2808}",
      journal = {\apj},
         year = 2015,
        month = jul,
       volume = {808},
       number = {1},
          eid = {51},
        pages = {51},
          doi = {10.1088/0004-637X/808/1/51},
archivePrefix = {arXiv},
       eprint = {1505.05934},
 primaryClass = {astro-ph.SR},
       adsurl = {https://ui.adsabs.harvard.edu/abs/2015ApJ...808...51M}
}

@ARTICLE{qiu2025,
       author = {{Qiu}, Tian and {Wang}, Wenting and {Koposov}, Sergey and {Li}, Ting S. and {Sandford}, Nathan R. and {Najita}, Joan and {Li}, Songting and {Han}, Jiaxin and {Dey}, Arjun and {Rockosi}, Constance and {Gaensicke}, Boris and {Han}, Jesse and {Weaver}, Benjamin Alan and {Myers}, Adam and {Aguilar}, Jessica Nicole and {Ahlen}, Steven and {Allende Prieto}, Carlos and {Bianchi}, Davide and {Brooks}, David and {Claybaugh}, Todd and {de la Macorra}, Axel and {Doel}, Peter and {Font-Ribera}, Andreu and {Forero-Romero}, Jaime and {Gaztanaga}, Enrique and {Gontcho}, Satya Gontcho A and {Gutierrez}, Gaston and {Jimenez}, Jorge and {Joyce}, Dick and {Kisner}, Theodore and {Lamman}, Claire and {Landriau}, Martin and {Le Guillou}, Laurent and {Meisner}, Aaron and {Miquel}, Ramon and {Nadathur}, Seshadri and {Percival}, Will and {Poppett}, Claire and {Prada}, Francisco and {Perez-Rafols}, Ignasi and {Rossi}, Graziano and {Sanchez}, Eusebio and {Schlegel}, David and {Silber}, Joseph Harry and {Sprayberry}, David and {Tarle}, Gregory and {Zhou}, Rongpu and {Zou}, Hu},
        title = "{The Binary Fraction of Stars in the Dwarf Galaxy Ursa Minor via Dark Energy Spectroscopic Instrument}",
      journal = {arXiv e-prints},
         year = 2025,
        month = dec,
          eid = {arXiv:2512.04477},
        pages = {arXiv:2512.04477},
          doi = {10.48550/arXiv.2512.04477},
archivePrefix = {arXiv},
       eprint = {2512.04477},
 primaryClass = {astro-ph.SR},
       adsurl = {https://ui.adsabs.harvard.edu/abs/2025arXiv251204477Q}
}

@ARTICLE{livernois2023,
       author = {{Livernois}, Alexander R. and {Vesperini}, Enrico and {Pavl{\'\i}k}, V{\'a}clav},
        title = "{Evolution of binary stars in the early evolutionary phases of ultra-faint dwarf galaxies}",
      journal = {\mnras},
         year = 2023,
        month = may,
       volume = {521},
       number = {3},
        pages = {4395-4405},
          doi = {10.1093/mnras/stad826},
archivePrefix = {arXiv},
       eprint = {2303.12841},
 primaryClass = {astro-ph.GA},
       adsurl = {https://ui.adsabs.harvard.edu/abs/2023MNRAS.521.4395L}
}

@ARTICLE{rieke2023,
       author = {{Rieke}, Marcia J. and {Kelly}, Douglas M. and {Misselt}, Karl and {Stansberry}, John and {Boyer}, Martha and {Beatty}, Thomas and {Egami}, Eiichi and {Florian}, Michael and {Greene}, Thomas P. and {Hainline}, Kevin and {Leisenring}, Jarron and {Roellig}, Thomas and {Schlawin}, Everett and {Sun}, Fengwu and {Tinnin}, Lee and {Williams}, Christina C. and {Willmer}, Christopher N.~A. and {Wilson}, Debra and {Clark}, Charles R. and {Rohrbach}, Scott and {Brooks}, Brian and {Canipe}, Alicia and {Correnti}, Matteo and {DiFelice}, Audrey and {Gennaro}, Mario and {Girard}, Julien H. and {Hartig}, George and {Hilbert}, Bryan and {Koekemoer}, Anton M. and {Nikolov}, Nikolay K. and {Pirzkal}, Norbert and {Rest}, Armin and {Robberto}, Massimo and {Sunnquist}, Ben and {Telfer}, Randal and {Wu}, Chi Rai and {Ferry}, Malcolm and {Lewis}, Dan and {Baum}, Stefi and {Beichman}, Charles and {Doyon}, Ren{\'e} and {Dressler}, Alan and {Eisenstein}, Daniel J. and {Ferrarese}, Laura and {Hodapp}, Klaus and {Horner}, Scott and {Jaffe}, Daniel T. and {Johnstone}, Doug and {Krist}, John and {Martin}, Peter and {McCarthy}, Donald W. and {Meyer}, Michael and {Rieke}, George H. and {Trauger}, John and {Young}, Erick T.},
        title = "{Performance of NIRCam on JWST in Flight}",
      journal = {\pasp},
         year = 2023,
        month = feb,
       volume = {135},
       number = {1044},
          eid = {028001},
        pages = {028001},
          doi = {10.1088/1538-3873/acac53},
archivePrefix = {arXiv},
       eprint = {2212.12069},
 primaryClass = {astro-ph.IM},
       adsurl = {https://ui.adsabs.harvard.edu/abs/2023PASP..135b8001R}
}

@INPROCEEDINGS{ford2003,
       author = {{Ford}, Holland C. and {Clampin}, Mark and {Hartig}, George F. and {Illingworth}, Garth D. and {Sirianni}, Marco and {Martel}, Andre R. and {Meurer}, Gerhardt R. and {McCann}, William J. and {Sullivan}, Pamela C. and {Bartko}, Frank and et al.},
        title = "{Overview of the Advanced Camera for Surveys on-orbit performance}",
    booktitle = {Future EUV/UV and Visible Space Astrophysics Missions and Instrumentation.},
         year = 2003,
       editor = {{Blades}, J. Chris and {Siegmund}, Oswald H.~W.},
       series = {Society of Photo-Optical Instrumentation Engineers (SPIE) Conference Series},
       volume = {4854},
        month = feb,
        pages = {81-94},
          doi = {10.1117/12.460040},
       adsurl = {https://ui.adsabs.harvard.edu/abs/2003SPIE.4854...81F}
}

@ARTICLE{raghavan2010R,
       author = {{Raghavan}, Deepak and {McAlister}, Harold A. and {Henry}, Todd J. and {Latham}, David W. and {Marcy}, Geoffrey W. and {Mason}, Brian D. and {Gies}, Douglas R. and {White}, Russel J. and {ten Brummelaar}, Theo A.},
        title = "{A Survey of Stellar Families: Multiplicity of Solar-type Stars}",
      journal = {\apjs},
         year = 2010,
        month = sep,
       volume = {190},
       number = {1},
        pages = {1-42},
          doi = {10.1088/0067-0049/190/1/1},
archivePrefix = {arXiv},
       eprint = {1007.0414},
 primaryClass = {astro-ph.SR},
       adsurl = {https://ui.adsabs.harvard.edu/abs/2010ApJS..190....1R}
}

\begin{appendix}
\label{app1}
\onecolumn

\section{On the age of Bo\"otes\,I}
\label{sec_age}
A visual inspection of the CMD shown in Fig.\,\ref{cmd0} reveals that the sparsely populated SGB of Bo\"otes\,I appears remarkably narrow and well defined. Since the SGB is highly sensitive to stellar age, in this Appendix we examine whether the current dataset allows us to place meaningful constraints on the ages of the stellar populations in Bo\"otes\,I. In addition, we assess how the choice of isochrone ages adopted throughout this paper affects the derived binary fractions and the inferred metallicity distribution.

\begin{figure*}[h]
    \centering
    \includegraphics[width=0.7\linewidth]{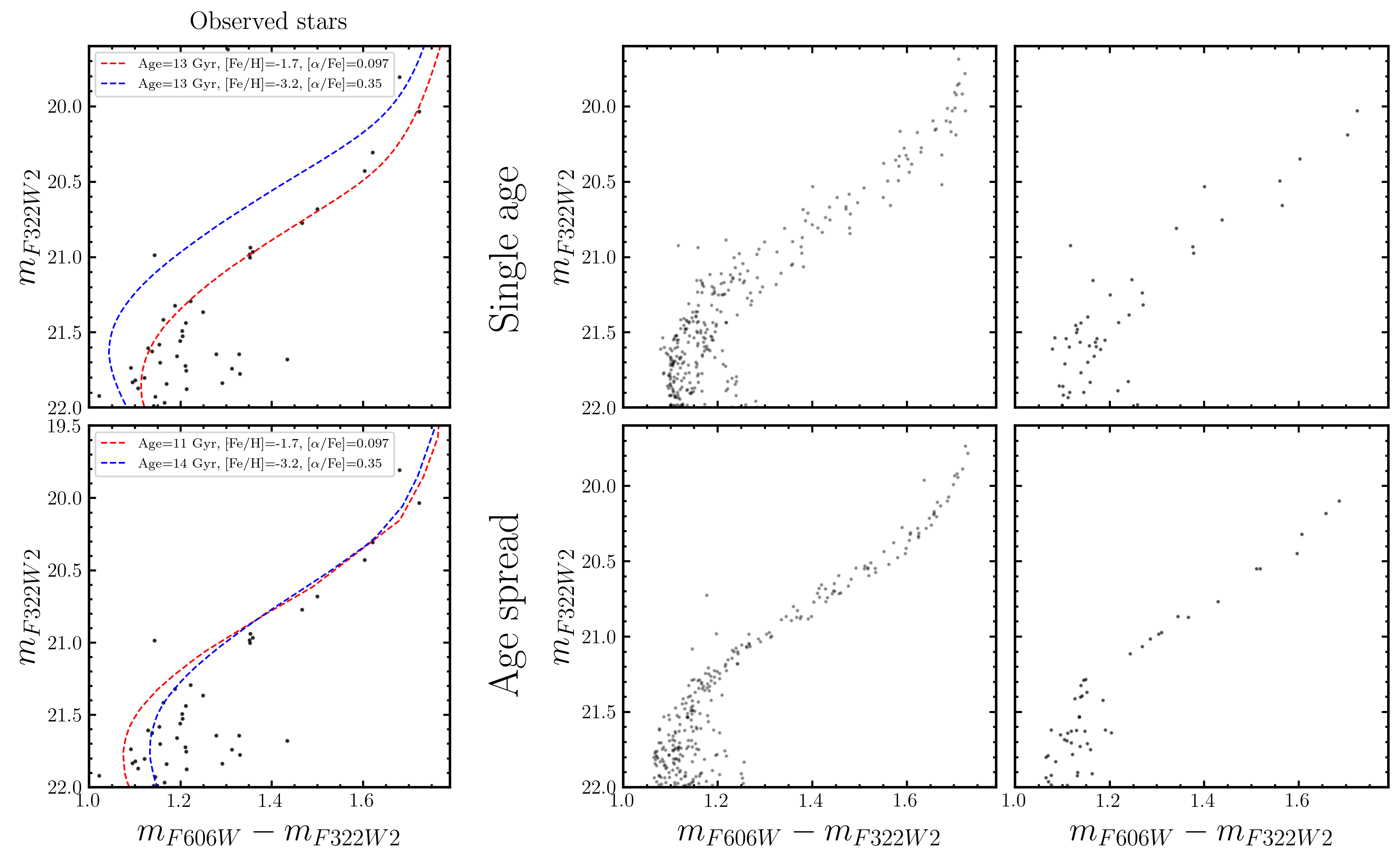}
    \caption{$m_{\rm F322W2}$ vs.\,$m_{\rm F606W}-m_{\rm F322W2}$ CMDs zoomed in around the SGB. 
The left panels show the observed CMDs of Bo\"otes~I, where the red and blue lines are isochrones with different [Fe/H], [$\alpha$/Fe], and ages, as indicated in the insets. 
The middle panels show simulated CMDs composed of coeval stellar populations (top) and of stellar populations with different ages (bottom). 
The right panels are the same as the middle panels, but include approximately the same number of stars as in the observed CMDs.}
    \label{cmd_age}
\end{figure*}

The left panels of Fig.\,\ref{cmd_age} reproduce the CMD shown in Fig.\,\ref{cmd0}, zoomed in around the SGB region, with two BaSTI isochrones overplotted, corresponding to the extreme metallicity values ([Fe/H]=$-3.2$ and $-1.7$). The bottom panel displays the isochrones with different ages (14 and 11 Gyr), introduced in Sect.,\ref{sec_met}, and adopted in this work. 

These two isochrones nearly overlap along the SGB.
The top panel, instead, shows isochrones of the same age (13 Gyr), which are widely separated in color along the SGB.

The middle panels of Fig.\,\ref{cmd_age} present two simulated CMDs. In both cases, we adopted the same luminosity function and binary fraction as in the observed CMD, but with a total number of stars increased by a factor of ten to reduce statistical noise. We also assumed the same metallicity distribution inferred in this study. The two simulated CMDs differ only in the age distribution of stellar populations with different metallicities.
The bottom-middle panel corresponds to the age distribution adopted in Sect.,\ref{sec_met}, whereas the top-middle panel assumes all populations are coeval. The right panels show the same simulated CMDs as in the middle panels, but rescaled to match the number of stars in the observed CMD.

Although the observed stars are distributed along a relatively narrow SGB---suggesting, at first glance, a prolonged star formation---the limited number of SGB stars prevents us from drawing firm conclusions about the ages of the Bo\"otes\,I stellar populations.
However, this hint of a prolonged star formation is in agreement with what was found by \cite{brown2014} and \cite{Durbin2025}, who inferred a primary star formation event about 13.4 Gyr ago, followed by residual star formation persisting until roughly 11 Gyr ago.

To assess the robustness of our results on the binary fraction and metallicity distribution, we repeated the analyses presented in Sects.\,\ref{sec_met} and \ref{sec_bin} using coeval isochrones. We verified that the results remain unchanged. This outcome is expected, since the determination of the binary fraction and metallicity distribution is based on MS stars, whose position in the CMD is largely insensitive to age.

\end{appendix}

\end{document}